\documentclass[floats,aps,superscriptaddress,amsfonts,notitlepage]{revtex4-1}
\usepackage{ifpdf}
\usepackage[utf8]{inputenc}
\usepackage[UKenglish]{babel}
\usepackage{graphicx}
\usepackage{hyperref}
\usepackage{amsmath, amsthm, amssymb}
\usepackage{multirow}
\usepackage{url}
\usepackage{subfigure}
\usepackage{color}
\usepackage{comment}

\newcommand{\cons}{{\text{cons}}}
\newcommand{\diss}{{\text{diss}}}
\newcommand{\en}{\mathcal{E}}
\newcommand{\ang}{\mathcal{L}}

\newcommand{\self}{\text{self}}
\newcommand{\full}{\text{full}}
\renewcommand{\max}{\text{max}}
\renewcommand{\min}{\text{min}}
\renewcommand{\c}{\cos \omega_r \tau}
\renewcommand{\b}{\bar}
\newcommand{\f}{\frac}

\newcommand{\nn}{\nonumber}

\newcommand{\el}{\ell}

\newcommand{\mrm}{\mathrm}
\newcommand{\be}{\begin{equation}}
\newcommand{\ee}{\end{equation}}
\newcommand{\ba}{\begin{eqnarray}}
\newcommand{\ea}{\end{eqnarray}}

\newcommand{\alert}[1]{#1}

\renewcommand{\c}{\hskip0.1cm,}
\newcommand{\p}{\hskip0.1cm.}

\mathchardef\mhyphen="2D

\def\prd{Phys. Rev. D}

\def\etal{\textit{et al.}}

\def\half{\frac{1}{2}}

\begin{document}

\title{Frequency-domain algorithm for the Lorenz-gauge gravitational self-force}
\author{Sarp Akcay}
\affiliation{School of Mathematics, University of Southampton, Southampton, SO17 1BJ, United Kingdom}
\author{Niels Warburton}
\affiliation{School of Mathematical Sciences and Complex \& Adaptive Systems Laboratory, University College Dublin, Belfield, Dublin 4, Ireland}
\author{Leor Barack}
\affiliation{School of Mathematics, University of Southampton, Southampton, SO17 1BJ, United Kingdom}

\begin{abstract}
State-of-the-art computations of the gravitational self-force (GSF) on massive particles in black hole spacetimes involve numerical evolution of the metric perturbation equations in the time-domain, which is computationally very costly. We present here a new strategy, based on a frequency-domain treatment of the perturbation equations, which offers considerable computational saving. The essential ingredients of our method are (i) a Fourier-harmonic decomposition of the Lorenz-gauge metric perturbation equations and a numerical solution of the resulting coupled set of ordinary equations with suitable boundary conditions; (ii) a generalized version of the method of extended homogeneous solutions [Phys. Rev. D {\bf 78}, 084021 (2008)] used to circumvent the Gibbs phenomenon that would otherwise hamper the convergence of the Fourier mode-sum at the particle's location; and (iii) standard mode-sum regularization, which finally yields the physical GSF as a sum over regularized modal contributions. We present a working code that implements this strategy to calculate the Lorenz-gauge GSF along eccentric geodesic orbits around a Schwarzschild black hole. The code is far more efficient than existing time-domain methods; the gain in computation speed (at a given precision) is about an order of magnitude at an eccentricity of 0.2, and up to three orders of magnitude for circular or nearly circular orbits. This increased efficiency was crucial in enabling the recently reported calculation of the long-term orbital evolution of an extreme mass ratio inspiral [Phys. Rev. D {\bf 85}, 061501(R) (2012)]. Here we provide full technical details of our method to complement the above report.

\end{abstract}	
\date{\today}
\maketitle

\section{Introduction}	

Astrophysical binaries of inspiralling compact objects are among the most promising sources for current and future gravitational-waves detector experiments. Their detection will offer insights into the fundamental workings of gravity in its most extreme regime.  The challenges associated with the detection and interpretation of such gravitational waves make it necessary to have at hand accurate theoretical models of the radiative dynamics in strongly interacting binaries. This need for precision models has produced a plethora of approaches to solving the relativistic two-body problem \cite{Blanchet:PN_review,Buonanno-Damour,Poisson-review,Baumgarte-Shapiro:NR_book}, each applicable in a particular domain of the problem.
When the  masses of the two components differ by orders of magnitude, the problem becomes amenable to perturbation theory in the small mass-ratio: At zeroth order the small object moves on a geodesic in the background spacetime of the larger one, and finite-mass corrections (due, e.g., to radiation reaction and internal structure) are accounted for, in principle, order by order in the mass ratio. In this effective description, the small object is subject to a gravitational self-force (GSF) exerted by its own gravitational field, with the latter thought of as a perturbation on the fixed geometry of the larger object.

The theoretical foundations for a robust formulation of the GSF in curved spacetime have been laid in the past decade and a half \cite{Mino-Sasaki-Tanaka,Quinn-Wald,Detweiler-Whiting,Gralla-Wald,Poisson-review}. Actual numerical calculations of the GSF for orbiting particles in black hole spacetimes have also been carried out \cite{Barack-Sago-circular,Detweiler-circular,Barack-Sago-eccentric,Akcay-GSF-circular}, often building on the techniques developed by considering scalar-field analogue models \cite{Detweiler-Messaritaki-Whiting, Burko-circular,Canizares-Sopuerta:circular,Canizares-Sopuerta:eccentric,Haas,Casals-Dolan-Ottewill-Wardell}. The state-of-the-art is a code that returns the GSF along any (fixed) bound geodesic orbit around a Schwarzschild black hole, and there are some preliminary attacks on the Kerr problem as well \cite{Shah_etal:Kerr,Dolan-Barack:Kerr_GSF}.
This success has lead to many fruitful exchanges with other approaches to the two-body problem \cite{LeTiec_etal-periastron_advance,Damour-EOB-SF,Barack-Damour-Sago,Blanchet-etal-PN-SF-comparision:circular,Akcay_etal}. For a review see \cite{Barack-review}.

Much of the work done so far has focused on calculating the GSF along a fixed geodesic orbit, without taking into account the back reaction on the orbit. A major priority task for the self-force program now is to devise a 
numerically efficient way of computing the orbital evolution under the full effect of the GSF. In principle, one could seek to solve the perturbation field equations and the self-forced equations of motion as a coupled set, in a self-consistent manner (as illustrated recently, using a scalar-field toy model, in Ref.\ \cite{Diener_etal:SSF_self_consistent}). However, such an approach would need to rely on computationally expensive time-domain methods.
Currently available time-domain codes run on large computer clusters, typically using hundreds of processors over a period of weeks, merely to compute a short inspiral of a few dozen orbits. Astrophysically relevant inspirals are expected to undergo hundreds of thousands of orbits whilst emitting gravitational waves at frequencies detectable by a LISA-like detector. Computing such long waveforms is a serious challenge for time-domain techniques, let alone the requirement to populate a template bank with tens of thousands of waveform templates.

An alternative approach, which is much more effective, is to construct an analytical model for the GSF by interpolating numerical GSF data computed along a (dense) sample of geodesic orbits. With such a model at hand, the orbital evolution can be computed quickly using the method of osculating elements \cite{Pound-Poisson:Osculating_orbits, Gair-etal}, in which the inspiral orbit is reconstructed from a smooth sequence of ``momentary'' tangent geodesics \footnote{In the relevant adiabatic scenario where the orbital evolution is slow, the error introduced by computing the GSF along fixed tangent geodesics, rather than along the true evolving orbit, should be of the same order as the error from neglecting second-order GSF effects. This formally justifies the use of ``geodesic'' GSF data at our order of approximation.}. It may still be required to produce large amounts of GSF data to inform the analytic fit, but this needs only be done once, after which any inspiral (starting with any initial conditions) may be computed at negligible computational cost. The main gain here comes from the fact that GSF data along bound geodesic orbits is relatively cheap to obtain, using frequency-domain (FD) methods. Previous work has demonstrated that FD codes can be faster than time-domain codes by orders of magnitude \cite{Barton-etal, Warburton-Barack:eccentric}, at least when the orbital eccentricity is not too large.  FD algorithms have become particularly efficient following the introduction of the method of extended homogeneous solutions (MEHS) \cite{Barack-Ori-Sago,Hopper-Evans:odd_sector}, which circumvents Gibbs-phenomenon complications arising from the finite differentiability of the perturbation field at the particle.

In a recent paper \cite{Warburton_etal} we reported a first computation of the orbital evolution, in the Schwarzschild case, using the above scheme of osculating elements with ``geodesic'' GSF input obtained via an analytic fit to FD data. The purpose of the current paper is to give full details of our FD method for computing the GSF. In particular, we will elaborate on the application of MEHS to the problem of calculating the GSF in Lorenz gauge (previously, MEHS has only been applied in calculations of the scalar-field self-force and the Regge-Wheeler-gauge GSF). This requires a generalization of MEHS to a set of coupled equations, which we describe here. The present paper may also be considered an extension of Ref.\ \cite{Akcay-GSF-circular}, where an FD-domain method has been applied to calculate the Lorenz-gauge GSF for {\it circular} orbits around a Schwarzschild black hole---here we generalize this to generic bound orbits.
Various technical aspects of that work, in particular the construction of the homogeneous radial fields and their boundary conditions, carry over to the calculation presented in this work. For that reason we shall refer to Ref.\ \cite{Akcay-GSF-circular} as Paper I, and, where appropriate, will refer the reader to it for further details.

The layout of this paper is as follows. In Sec.~\ref{sec:prelim} we review relevant background material concerning the parametrization and characteristics of bound geodesic orbits in Schwarzschild geometry, 
write down the corresponding sourced Lorenz-gauge perturbation equations, and give a FD reformulation of these equations.
In Sec.~\ref{sec:MEHS_coupled_fields} we generalize MEHS to the case of coupled fields, relevant for our Lorenz-gauge analysis.
Section \ref{sec:GSF} reviews relevant results in GSF physics, and in particular the mode-sum approach used in our work. Section \ref{sec:numerical_implemenation} gives an algorithmic description of our numerical method, from the construction of physical boundary conditions for the FD metric perturbation, to the reconstruction of the GSF from a sum over Fourier-harmonic modes. In Sec.\ \ref{sec:nearly_static_modes} we describe a certain problem that hinders our computation when very low frequency modes are encountered, and propose a mitigation method. We present a sample of numerical results in Sec.~\ref{sec:results}, and in Sec.\ \ref{Sec:outlook} give an outlook of foreseeable extensions of our method.

Throughout this work we use geometrized units such that the speed of light and the gravitational constant are equal to unity. We shall denote the mass of the background Schwarzschild geometry by $M$ and the mass of the orbiting particle by $\mu$, with the assumption $\mu/M\ll 1$. We use metric signature $(-+++)$.

\section{Preliminaries: Field equations and Fourier-harmonic decomposition }\label{sec:prelim}

\subsection{Orbital parametrization}\label{sec:Schwarzschild_geodesics}

We start by reviewing bound geodesic orbits in Schwarzschild geometry. We shall denote the worldline of the test body (the point-mass particle) by $x^\alpha=x^\alpha_p(\tau)$ and its tangent four velocity by $u^\alpha = dx_p^\alpha/d\tau$, where $\tau$ is the body's proper time. In the geodesic approximation, the motion of the test body is governed by 
\begin{equation}\label{eq:geodesic_motion}
	\mu u^\beta\nabla_\beta u^\alpha = 0 ,
\end{equation}
where the covariant derivative is taken with respect to the background (Schwarzschild) metric $g$.  Using Schwarzschild coordinates, Eq.\ \eqref{eq:geodesic_motion} can be written explicitly as
\begin{eqnarray}
	\frac{dt_p}{d\tau} = \frac{\en}{f(r_p)}\c	&\;&\qquad \frac{d\varphi_p}{d\tau} = \frac{\ang}{r^2_p}\c	\\
	\left(\frac{dr_p}{d\tau}\right)^2 = \en^2 - R(r_p;\ang^2)\,, &\;&\qquad R(r;\ang^2) \equiv f(r)\left(1+ \frac{\ang^2}{r^2} \right)	\label{eq:R_effective}\c
\end{eqnarray}
where $\en\equiv-u_t$ and $\ang\equiv u_\varphi$ are the integrals of motion corresponding to the test body's specific energy and angular momentum, respectively, $f(r)\equiv 1 - 2M/r$ and $R(r;\ang^2)$ is an effective potential for the radial motion. In this work we shall be concerned solely with {\it bound} geodesic motion and so we specialize immediately to this case. Such orbits are specified uniquely, up to initial phase, by their energy and angular momentum, with $\frac{2\sqrt{2}}{3}<\en<1$ and $\ang > 2\sqrt{3}M$. 

Following Newtonian celestial mechanics, it will be useful to introduce an alternative, more geometrically motivated, orbital parametrization given by the dimensionless semi-latus rectum, $p$, and orbital eccentricity, $e$. Let the libration region be given by $r_\min\leq r_p\leq r_\max$, with $r_\min$ and $r_\max$ being the periastron and apastron radii.  Then $p$ and $e$ are defined through
\begin{eqnarray}
	p \equiv \frac{2\:r_{\max}\: r_{\min}}{M(r_{\max} + r_{\min})}\,,	\quad	e \equiv \frac{r_{\max} - r_{\min}}{r_{\max} + r_{\min}}	\label{eq:p_e_definition}\p
\end{eqnarray}
Note that $e=0$ for circular orbits (when $r_{\max}=r_{\min}$) and $e\rightarrow 1$ as $r_{\max}\rightarrow \infty$ (with fixed $r_{\min}$). Thus we have $0\le e < 1$. The range of $p$ will be constrained below. Equations \eqref{eq:p_e_definition} can be inverted to give $r_{\max}$ and $r_{\min}$ in terms of $p$ and $e$:
\begin{eqnarray}\label{eq:r1_and_r2}
	r_{\max} = \frac{pM}{1-e}\,, \quad r_{\min} = \frac{pM}{1+e}	\p
\end{eqnarray}
The (one-to-one) relation between $(p,e)$ and $(\en, \ang)$ is given by
\begin{eqnarray}\label{eq:Schwarzschild_Energy_Ang_Mom}
	\en^2 = \frac{(p-2-2e)(p-2+2e)}{p(p-3-e^2)}\,, \quad \ang^2 = \frac{p^2 M^2}{p-3-e^2}	\p
\end{eqnarray}
Again in a analogy with Newtonian celestial mechanics, and following Darwin \cite{Darwin-1961}, we introduce a ``relativistic anomaly'' parameter $\chi$, such that the radial motion is given by
\begin{eqnarray}\label{eq:r_p}
	r_p(\chi) = \frac{pM}{1+e \cos\chi}	\p
\end{eqnarray}
$\chi=0$ and $\chi=\pi$ correspond to periastron and apastron passages, respectively. Each of the parameters $t_p$, $\varphi_p$ and $\chi$ is monotonically increasing along the orbit;  the relation between these parameters is given by \cite{Cutler-Kennefick-Poisson}
\begin{eqnarray}
	\frac{dt_p}{d\chi} &=& \frac{Mp^2}{(p-2-2e\cos\chi)(1+e\cos\chi)^2}\sqrt{\frac{(p-2-2e)(p-2+2e)}{p-6-2e\cos\chi}}	\label{eq:dt_dchi}\c		\\
	\frac{d\varphi_p}{d\chi} &=& \sqrt{\frac{p}{p-6-2e\cos\chi}}	\label{eq:dphi_dchi}	\p
\end{eqnarray}
Without loss of generality we shall assume $t_p=\varphi_p=0$ at $\chi=0$. 

The accumulated azimuthal angle over one radial orbit (between two successive periastra) is found to be
\begin{eqnarray}\label{eq:Delta_phi}
	\Delta\varphi = \int_0^{2\pi} \frac{d\varphi_p}{d\chi}\,d\chi = 4 \left(\frac{p}{p-6-2e} \right)^{1/2}K\left(\frac{4e}{p-6-2e}\right)		\c
\end{eqnarray}
where $K(k)=\int_0^{\pi/2} (1-k\sin\theta)^{-1/2}\,d\theta$ is the complete elliptic integral of the first kind. We have $\Delta\varphi\ne 2\pi$, and the orbit precesses. The radial period, in terms of Schwarzschild time $t$, is given by
\begin{eqnarray}\label{eq:T_r}
	T_r = \int_0^{2\pi} \frac{dt_p}{d\chi}\,d\chi		\c
\end{eqnarray}
and the corresponding radial and (average) azimuthal frequencies are given by
\begin{eqnarray}
	\Omega_r = \frac{2\pi}{T_r}\,, \qquad \Omega_\varphi = \frac{\Delta\phi}{T_r}	\p \label{eq:frequencies}
\end{eqnarray}

In the case of circular orbits $(e=0,p\equiv r_0/M)$, the above orbital frequencies reduce to
\begin{eqnarray}\label{eq:Schwarzschild_orbital_freqs}
	\Omega_r^0 = \sqrt{\frac{(r_0-6M)M}{r_0^4}}		\c 	\qquad \Omega_\varphi^0 = \left(\frac{M}{r^{3}}\right)^{1/2}\c		\label{eq:circular_orbit_freqs}
\end{eqnarray}
where hereafter a sub/superscript `0' denotes the circular-orbit limit of a quantity. The $\Omega_r$ frequency of circular orbits is identified with the radial frequency of an infinitesimal eccentricity perturbation. Circular orbits with $r_0 > 6M$ are stable to eccentricity perturbations, whilst orbits with $3M < r_0 < 6M$ are unstable. At $r_0=3M$ only massless particles can orbit the black hole and this $r_0$ value is said to be the radius of the {\it light ring}. Below the light ring there are no circular timelike or null geodesics, stable or unstable. The circular orbit with radius $r_0=6M$ is known as the innermost stable circular orbit (ISCO).

\begin{figure}
	\centering
	\includegraphics[width=8cm]{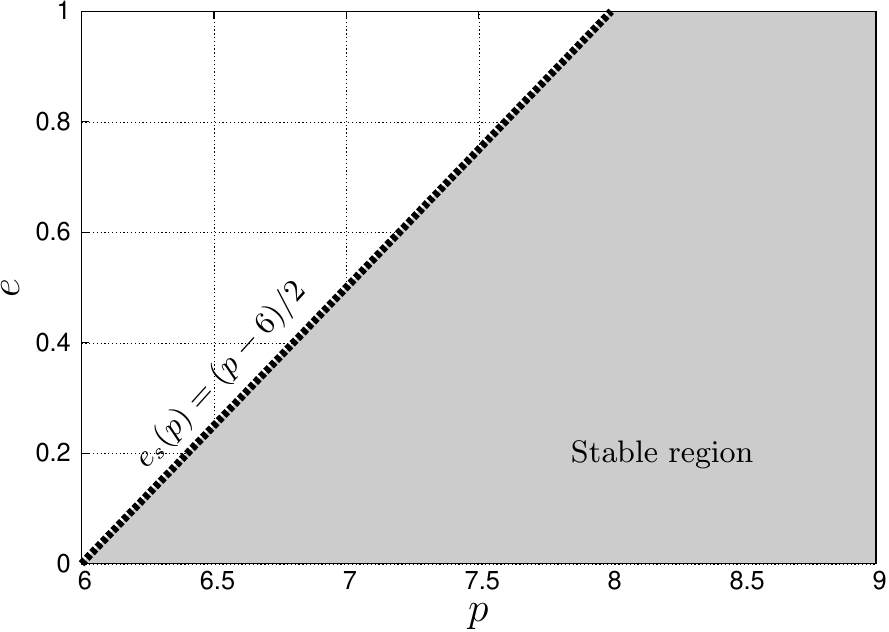}
	\caption[The $(p,e)$ orbital parameter space]{The $(p,e)$ orbital parameter space. The shaded region bounded by $0\le e\le1$ and $6+2e \le p < \infty$ marks the space of bound stable orbits. Circular orbits correspond to $e=0$, and the intersection of the separatrix $e_s(p)=(p-6)/2$ with the line $e=0$ marks the location of the innermost stable circular orbit (ISCO).}\label{fig:pe_space}
\end{figure}

For eccentric orbits, there exists a separatrix in the $(p,e)$ parameter space between the space of bound stable orbits and the space of unstable orbits. The value of $p$ at the separatrix is given by the curve $ p = 6+2e $ \cite{Cutler-Kennefick-Poisson} .
We plot the region of stable and unstable orbits in Fig. \ref{fig:pe_space}. For orbits along the separatrix, both $\Delta\varphi$ and $T_r$ diverge (while $\Omega_\varphi$ remains finite). This is a manifestation of the well known zoom-whirl behavior of near-separatrix orbits, where the particle can orbit (`whirl') many times just outside the periastron radius before `zooming' back out to the apastron \cite{Glampedakis-Kennefick}.

\subsection{Lorenz-gauge perturbation equations and multipole decomposition}\label{sec:gravitational_pert_decomp}

We proceed to give an overview of the Lorenz-gauge perturbation equations and their decomposition in Schwarzschild spacetime into tensor spherical harmonics. We follow the notation of Barack and Lousto \cite{Barack-Lousto-2005}, and refer the reader to that article for further details. 

Let us denote the full spacetime metric by $\mathfrak{g}$, which we shall consider to be the sum of the metric perturbation, $h$, and the background Schwarzschild metric, $g$. We thus have  $\mathfrak{g} = g + h$.  Linearizing the Einstein field equations in $h$ about $g$ yields the perturbation equations 
\begin{eqnarray}\label{eq:linearized_einstein}
	\square\bar{h}_{\mu\nu} + 2 R^{\alpha\;\;\beta}_{\;\;\mu\;\;\nu}\bar{h}_{\alpha\beta} = -16\pi T_{\mu\nu}	\c
\end{eqnarray}
where $\square = \nabla_\mu\nabla^\mu$ (with $\nabla_\mu$ denoting covariant differentiation with respect to $g$ ), $R$ is the Riemann tensor associated with $g$,
\begin{equation}
	\bar{h}_{\mu\nu} \equiv h_{\mu\nu} - \frac{1}{2}g_{\mu\nu}g^{\alpha\beta}h_{\alpha\beta}
\end{equation}
is the ``trace-reversed'' metric perturbation, and we have imposed the Lorenz gauge condition 
\begin{equation}\label{eq:lorenz_gauge}
	\nabla_\mu \bar{h}^{\mu\nu}=0 \p
\end{equation}
In this work the metric perturbation is sourced by a point particle of mass $\mu$, with energy-momentum 
\begin{equation}\label{eq:GSF_source}
	T_{\mu\nu}(x^\alpha) = \mu\int^\infty_{-\infty} [-\det(g)]^{-1/2} \delta^4(x^\alpha - x_p^\alpha)u_\mu u_\nu\, d\tau		\p
\end{equation}
The gauge equation \eqref{eq:lorenz_gauge} and field equation \eqref{eq:linearized_einstein} are consistent so long as the particle is moving on a geodesic of the background spacetime (as then $\nabla_\mu T^{\mu\nu} = 0$). 

The field equation \eqref{eq:linearized_einstein} is not easily amenable to direct numerical treatment, since its physical (retarded) solutions are singular at the particle 
(see, however, Refs.~\cite{Barack-Golbourn, Vega-Detweiler,Vega-Wardell-Diener,Dolan-Barack:GSF_m_mode} for techniques to overcomes this problem). In this work we choose to decompose the metric perturbation into  (tensorial) spherical harmonic modes. A key motivation is that the individual modes are everywhere bounded and continuous thus easier to work with. We then further decompose the multipoles into Fourier modes (as described in the next subsection), reducing the system to a set of ordinary differential equations (ODEs). 

The decomposition of the metric perturbation into multipole modes is achieved by projecting $\bar{h}_{\mu\nu}$ onto a basis of 2nd-rank tensor harmonics, defined (in the background Schwarzschild geometry) on 2-spheres with $t,r=\text{const}$. The spherical symmetry of the background geometry ensures that the individual multipole harmonics are eigenfunctions of the wave operator on the left-hand side of Eq.\ \eqref{eq:linearized_einstein}. The individual multipole modes hence decouple, though, in general, the ten tensorial components of each multipole mode will remain coupled.

We shall use here the tensorial-harmonic basis $Y_{\mu\nu}^{(i)lm}(\theta,\varphi;r)$ (where $i=1,\ldots,10$) defined in \cite{Barack-Sago-circular}. [The definition involves certain multiplicative factors of $r$ and $f(r)$, introduced in order to balance the dimensions and simplify the resulting equations.]
The $Y_{\mu\nu}^{(i)lm}$s form a basis for any second rank, symmetric tensor field in 4 dimensions. They are orthonormal in the sense that
\begin{equation}
	\int \eta^{\alpha\mu}\eta^{\beta\nu}\left(Y_{\mu\nu}^{(i)lm}\right)^* Y_{\alpha\beta}^{(j)l'm'}\,d\Omega = \delta_{ij}\delta_{ll'}\delta_{mm'}\c
\end{equation}
where  $\eta^{\alpha\mu}=\text{diag}(1,f^2,r^{-2},r^{-2}\sin^{-2}\theta)$, an asterisk denotes complex conjugation, and $ d\Omega = \sin\theta \: d\theta \: d\phi $. 

We expand the energy-momentum tensor in Eq.~\eqref{eq:GSF_source} in the form
\begin{equation}
	T_{\mu\nu} = \sum_{lm}\sum_{i=1}^{10} T^{(i)}_{lm}(t,r)Y^{(i)lm}_{\mu\nu}\c
\end{equation}
where the harmonic coefficients are given by
\begin{eqnarray}
	T^{(i)}_{lm}(t,r) 	&=& \int d\Omega\, T_{\mu\nu}\eta^{\mu\alpha}\eta^{\nu\beta}\left(Y_{\alpha\beta}^{(i)lm}\right)^*	\\
	&=& \frac{\mu}{u^t r_p^2}u_\mu u_\nu \: \eta^{\mu\alpha}(x_p) \:\eta^{\nu\beta}(x_p)\: \left(Y_{\alpha\beta}^{(i)lm}(\theta_p,\varphi_p;r_p)\right)^*\delta(r-r_p)\p
\end{eqnarray}
We similarly expand the metric perturbation in the form
\begin{eqnarray}
	\bar{h}_{\mu\nu}(t,r,\theta,\phi) = \frac{\mu}{r}\sum_{lm}\sum_{i=1}^{10}\bar{h}^{(i)lm}(t,r)Y_{\mu\nu}^{(i)lm}(\theta,\varphi;r)		\c
\end{eqnarray}
and substitute it into the linearized Einstein equation \eqref{eq:linearized_einstein}, whereupon individual $l,m$-modes decouple and the angular dependence separates out of the equations. The time-radial scalar-like functions $\bar{h}^{(i)lm}(t,r)$ (numbering 10 for each $l,m$) obey the coupled set of partial differential equations
\begin{eqnarray}\label{eq:metric_pert_1+1}
	\square^{sc}_l \bar{h}_{lm}^{(i)} + \mathcal{M}^{(i)l}_{(j)} \bar{h}_{lm}^{(j)} = 4\pi\mu^{-1}rfT^{(i)lm} \equiv \mathcal{S}_{lm}^{(i)}\delta(r-r_p) \quad (i=1,...,10)\c
\end{eqnarray}
where $\square^{sc}_l$ is the scalar-field wave operator,
\begin{eqnarray}
	\square^{sc}_l = \frac{1}{4}\left[\partial_t^2 - \partial_{r_*}^2 + V_l(r)\right]	\c
\end{eqnarray}
with the potential term given by
\begin{equation}
	V_{l}(r) = f(r)\left[\frac{2M}{r^3} + \frac{l(l+1)}{r^2}\right]	\p
\end{equation}
We have also introduced the standard tortoise coordinate, $r_*$, defined by $dr_*/dr=f(r)^{-1}$, giving
\begin{equation}\label{eq:tortoise_coord}
	r_* = r + 2M\ln\left(\frac{r}{2M}-1\right)		\c
\end{equation}
where we have specified the constant of integration. The terms $\mathcal{M}^{(i)l}_{(j)}$ appearing in Eq.\ \eqref{eq:metric_pert_1+1} are first-order differential operators that couple between the ten components of the metric perturbation. The explicit form of $\mathcal{M}^{(i)l}_{(j)}$ can be found in Appendix B of Ref.~\cite{Barack-Sago-eccentric}. We give the source coefficients $\mathcal{S}^{(i)}$ and FD versions of $\mathcal{M}^{(i)l}_{(j)}$ in Appendix \ref{apdx:field_eqs_and_sources}. 

It will be useful to note that the ten field equations \eqref{eq:metric_pert_1+1} are not all coupled together, but form two disjoint sets of equations, one for each parity: Basis elements with $i=1,\dots,7$ have {\it even} parity, remaining unchanged under the parity operation $(\theta,\varphi) \rightarrow (\pi-\theta,\varphi+\pi)$. Basis elements $i=8,9,10$ change sign under parity and hence are {\it odd}. For {\em equatorial} orbits one finds
\begin{align}
	\mathcal{S}^{(i=1,\dots,7)} &\propto \quad \left[Y^{lm} (\pi/2,\varphi_p) \right]^\ast= 0\qquad \text{for}\qquad l+m = \text{odd}		\c		\\
	\mathcal{S}^{(i=8,9,10)} &\propto \left[\partial_\theta \: Y^{lm}(\theta,\varphi_p)\right]^\ast_{\theta=\pi/2} = 0\qquad \text{for}\quad  l+m = \text{even}	\p
\end{align}
As a result, 
$\bar{h}^{(i=1,\dots,7)}$ vanish trivially for $l+m=\text{odd}$, and $\bar{h}^{(i=8,9,10)} $ vanish trivially for $l+m=\text{even}$.

\subsection{Fourier decomposition}\label{sec:FD_decomp}

At this point we depart from the 1+1D treatment of Refs.\ \cite{Barack-Sago-circular,Barack-Sago-eccentric}, and introduce a decomposition of the field equations into (Fourier) frequency modes.
For bound geodesic orbits, the spectrum of the Fourier decomposition is found to be discrete (see, e.g., Appendix D.2 of Ref.~\cite{Barack-Ori-Sago}), with each mode labelled by two integers---the azimuthal number, $m$, and the Fourier number, $n$. The mode frequency is given by 
\begin{equation}
\omega = m\Omega_\varphi + n \Omega_r \c
\end{equation}
where $\Omega_r$ and $\Omega_\varphi$ are the orbit frequencies given in Eq.\ \eqref{eq:frequencies}. We can, therefore, write the $(t,r)$ dependence of the trace reversed metric perturbation as a sum over discrete Fourier modes,
\begin{equation}
	\bar{h}^{(i)}_{lm}(t,r) = \sum_n R^{(i)}_{lmn}(r)e^{-i\omega t}\p \label{eq:Fourier_decomp}
\end{equation}
By substituting the above into Eq.~\eqref{eq:metric_pert_1+1}, one finds that the radial dependence of the (trace-reversed) metric perturbation completely separates, and the field equations reduce to a set of 10 coupled ODEs (one set for each $l,m,n$):
\begin{equation}\label{eq:metric_pert_FD}
	\frac{d^2}{dr_*^2} R^{(i)}_{lmn}(r) - \left[V_{l}(r) - \omega^2 \right] R^{(i)}_{lmn}(r) - 4\hat{\mathcal{M}}^{(i)l}_{(j)}R^{(j)}_{lmn}(r) =  J^{(i)}_{lmn}	\c 
\end{equation}
where $\hat{\mathcal{M}}^{(i)l}_{(j)}$ are the Fourier-transformed versions of $\mathcal{M}^{(i)l}_{(j)}$, and $J^{(i)}_{lmn}$ are related to the Fourier-transforms of $\mathcal{S}^{(i)}_{lm}$. The explicit form of $\hat{\mathcal{M}}^{(i)l}_{(j)}$ and the source terms $ J^{(i)}_{l m n}$ are presented in Appendix \ref{apdx:field_eqs_and_sources} (generalizing the source terms of Paper I to eccentric orbits). We note that the separation under parity of the 1+1D field equations carries over to the FD. Thus $R^{(i=1,\dots,7)} = 0$ for $l+m=\text{odd}$ and $R^{(i=8,9,10)} = 0$ for $l+m=\text{even}$.

The Fourier-harmonic decomposition of the Lorenz-gauge condition $\nabla_\mu \bar{h}^\mu_{\,\,\,\nu} = 0$ results in four equations that also separate under parity, with the first three (as ordered below) involving only even-parity modes and the fourth involving only odd-parity modes. For each $lmn$-mode these equations read
\begin{gather}
	i\omega R^{(1)} + f\left(i\omega R^{(3)} + R^{(2)}_{,r} + \frac{R^{(2)}}{r} - \frac{R^{(4)}}{r}\right) = 0		\c				\label{eq:gauge1}	\\
	-i\omega R^{(2)} - fR^{(1)}_{,r} + f^2 R^{(3)}_{,r} - \frac{f}{r}\left( R^{(1)} - R^{(5)} - fR^{(3)} - 2fR^{(6)} \right) = 0\c	\label{eq:gauge2}	\\
	-i\omega R^{(4)} - \frac{f}{r}\left(r R^{(5)}_{,r} + 2R^{(5)} + l(l+1)R^{(6)} - R^{(7)} \right) = 0			\c					\label{eq:gauge3}	\\
	-i\omega R^{(8)} - \frac{f}{r}\left(r R^{(9)}_{,r} + 2R^{(9)} - R^{(10)} \right) = 0			\p								\label{eq:gauge4}
\end{gather}
where for brevity we have dropped the $lmn$ indices.

\subsubsection{Hierarchical structure of the FD field equations}

The gauge conditions can be used to reduce the number of field equations that need to be solved for simultaneously. Odd-parity modes have three coupled fields ($i=8,9,10$) in general, but we can use the gauge condition \eqref{eq:gauge4} (when $\omega\ne 0$) to obtain $R^{(8)}$ algebraically in terms of $R^{(9)}$ and $R^{(10)}$. Since $R^{(8)}$ does not feature in the field equations for $R^{(9)}$ and $R^{(10)}$, we may first solve the coupled set for the latter two, and then obtain $R^{(8)}$ from the gauge condition.  Hence, in the odd sector we face solving a coupled set of two equations only. In the special case $(l,m)=(1,0)$ the function $R^{(10)}$ vanishes trivially, and the system reduces further to a single field equation for $R^{(9)}$.

The even-parity sector consists of seven coupled fields ($R^{(1)},\ldots,R^{(7)}$) and three gauge constraint equations [Eqs.~\eqref{eq:gauge1}--\eqref{eq:gauge3}]. In principle, thus, we need only solve for four radial fields and we may then construct the remaining three fields using the gauge equations. In practice, however, it is simpler [following Ref.~\cite{Barack-Sago-eccentric}] to solve the field equations for the {\em five} coupled fields $R^{(i)}$ with $i=1,3,5,6,7$, and then make use of Eqs.~\eqref{eq:gauge2} and \eqref{eq:gauge3} to obtain $R^{(2)}$ and $R^{(4)}$ (neither of the last two functions feature in the set of five coupled field equations). We can use the remaining gauge equation \eqref{eq:gauge1} as a consistency check on our numerical results. In the special case $(l,m)=(1,\pm 1)$ the function $R^{(7)}$ vanishes trivially, and the system of field equations reduces to four equations for $i=1,3,5,6$. For $(l,m)=(0,0)$ both $R^{(7)}$ and $R^{(5)}$ vanish, and the system further reduces to just three coupled equations, for $i=1,3,6$.

The static mode $m=0=n$ (for each $l$) is dealt with differently. In this case, the functions $R^{(9)}$ and $R^{(10)}$ (in the odd-parity sector) and $R^{(2)}$ and $R^{(4)}$ (in the even-parity sector) vanish identically, and the above structure changes. The gauge equation \eqref{eq:gauge4} becomes trivial [as does \eqref{eq:gauge1}] and it cannot be used to obtain $R^{(8)}$; instead, we obtain this function by solving the $i=8$ field equation. In the even sector, the remaining gauge conditions, \eqref{eq:gauge2} and \eqref{eq:gauge3}, can be used to express $R^{(6)}$ and $R^{(7)}$ algebraically in terms of $R^{(1)}$, $R^{(3)}$ and $R^{(5)}$ (and their first derivatives), which in turn can be used to decouple the subset of field equations with $i=1,3,5$ from the rest of the set. One thus solves for $i=1,3,5$ and then obtains $i=6,7$ using these algebraic expressions. 

The above hierarchical scheme for constructing the fields $R^{(i)}$ is summarized in Table \ref{table:hierarchical_structure}. The table shows the variety of different cases, depending on the values of $lmn$.

\begin{table}
\centering
\begin{tabular} {| c | c || l   ||  l | }
    \hline \hline
    \multicolumn{2}{|c||}{case}& $l+m=\text{even}$ &  $l+m=\text{odd} $ \\
    \hline \hline
    $ l \ge 2 $ & $ |m|+|n|\ne 0 $ &$ i = 1,3,5,6,7 \rightarrow 2,4 $ &  $i =9, 10 \rightarrow 8 $  \\
	 \: & $ m=0=n $ & $ i=1, 3, 5 \rightarrow 6,7 $ &   $i=8 $ only  \\    
	 \hline
	 $l=0,\ m=0 $ & $ n\ne 0 $ & $i = 1, 3, 6 \rightarrow 2 $  &  \quad\quad --  \\
	 \ & $n=0$ & $ i = 1, 3 \rightarrow 6 $ &   \quad\quad --   \\
	 \hline
	  $ l = 1,\ m=0 $ & $n\ne 0$ & \quad\quad  -- &  $i = 9 \rightarrow 8$  \\
	   & $n=0$ &  \quad\quad  -- &  $i=8$ only  \\
	   \hline
	  $ l = 1,\ m=\pm 1 $ &  & $ i=1, 3, 5, 6 \rightarrow 2, 4 $ & \quad\quad  --  \\
    \hline \hline
    \end{tabular}
\caption{Hierarchical scheme for solving the coupled ODEs (\ref{eq:metric_pert_FD}) for $R^{(i)}_{lmn}(r)$ in each case. Arrows (`$\rightarrow$') indicate algebraic construction using the gauge equations \eqref{eq:gauge1}-\eqref{eq:gauge4}. In general, all tensorial components $i=1,\ldots,10$ are excited (first row of table), but there are some special modes, specifically the static $m=0=n$ modes and the low multipoles $l=0,1$, for which some of the tensorial components vanish identically. These special modes are displayed separately in the table. ``Resonant'' static modes with $\omega=0$ but nonzero $m,n$ (when they occur) require a separate treatment; we discuss these modes briefly in Sec.\ \ref{sec:low_freq_modes} and in more detail in Sec.\ \ref{sec:nearly_static_modes}. 
}\label{table:hierarchical_structure}
\end{table}

\subsubsection{Low-frequency modes}\label{sec:low_freq_modes}

We have mentioned the special mode $m=n=0$, which is static (i.e., has $\omega=0$). One such mode must be calculated for each value of $l$. Our numerical algorithm may encounter yet another type of static mode, for which $m,n\ne 0$ but $\omega=m\Omega_\phi+n\Omega_r=0$. Such ``resonant'' modes \footnote{Our ``resonant'' modes should not be confused with the phenomenon first discussed by Flanagan and Hinderer in Ref.~\cite{Hinderer-Flanagan:resonances}. The latter are true dynamical resonances occurring between the two intrinsic (libration-type) frequencies of generic orbits around a Kerr spacetime.} occur when the frequency ratio $\Omega_\phi/\Omega_r$ happens to be a rational number $-n/m$ where $n$ and $m$ are the indices of the modes that need to be calculated numerically. (In an actual numerical implementation one always truncates the mode sum at some finite values of $n$ and $m$, so only certain ``low-order'' resonances are relevant in practice.) Resonant modes require a special treatment, both in the formulation of boundary conditions (see the discussion in Sec.\ \ref{sec:BCs}) and because the standard basis of homogeneous solutions degenerates in the resonant case (see Sec.\ \ref{sec:nearly_static_modes}).

Orbits for which there exist (low order) modes that are {\it precisely} resonant constitute a set of measure zero in the parameter space, so they may be avoided in certain applications. However, as we discuss in Sec.\ \ref{sec:nearly_static_modes}, a substantial portion of the parameter space is covered by orbits for which there occur {\it nearly} resonant modes: ones with $M|\omega|$ values small enough to cause numerical difficulties. In fact, all orbits of sufficiently large $p$ are ``near-resonant'' with respect to $(m,n)=(\pm 1,\mp 1)$, since the difference $\Omega_\varphi-\Omega_r$ decays rapidly (like $\sim 3r_p^{-5/2}$) at large $r_p$. Low-frequency modes prove difficult to deal with using our numerical method. In Sec.\ \ref{sec:nearly_static_modes} we will discuss this problem in more detail and propose a way to mitigate it. In the meantime, through Secs.\ \ref{sec:MEHS_coupled_fields}--\ref{sec:numerical_implemenation}, we ignore this issue.

\subsection{Physical boundary conditions}\label{sec:BCs}
The FD field equations (\ref{eq:metric_pert_FD}) must be solved subject to appropriate boundary conditions at $r_*\to \pm\infty$. Since we are interested in constructing the physical, retarded solutions, non-stationary modes (ones with $\omega\ne 0$) should represent purely outgoing radiation at infinity, $r_*\to +\infty$, and purely ingoing radiation at the horizon, $r_*\to -\infty$. At the level of the time-domain fields $\bar{h}^{(i)}$, the conditions are
\be
\bar{h}^{(i)}\sim \mrm{e}^{- i \omega (t \mp r_\ast) }, \label{eq:physical_BC}
\ee
where the upper sign corresponds to future null infinity, and the lower sign corresponds to the future event horizon. From this we can read
\be
R^{(i)}(r_*\to \pm\infty) \sim \mrm{e}^ {\pm i \omega r_\ast },
 \label{eq:physical_BC_in_FD}
\ee
for any $l,m,(i)$ and any $\omega\ne 0$. For $\omega=0$ modes these conditions are replaced with the requirement that the radial solutions are regular functions at $r=\infty$ and at $r=2M$.

In practice we will be solving the field equations numerically, and we cannot place the inner and outer boundaries of our numerical domain at $ r_\ast = \pm \infty $. Instead we will devise approximate boundary conditions at finite (large) values of $|r_*|$. This will be described in Sec.\ \ref{sec:numerical_BC}.

\section{Construction of the inhomogeneous fields: method of extended homogeneous solutions for coupled fields}\label{sec:MEHS_coupled_fields}

The calculation of the GSF via mode-sum regularization (see the next section) involves the construction of the time-domain fields $\bar h_{lm}^{(i)}(t,r)$ and their first derivatives at the location of the particle. In the standard FD approach, these values are to be obtained from an (inverse) Fourier sum over frequency modes $R_{lmn}^{(i)}(r)$, which are solutions to the inhomogeneous equation \ (\ref{eq:metric_pert_FD}). In general, the time-domain modes $\bar h_{lm}^{(i)}(t,r)$, thought of as functions of $t$ at fixed $r$, are non-smooth across the particle's worldline---their $t$ derivatives are generally discontinuous there (unless the orbit is circular, or one evaluates the derivatives at a radial turning point). This means that an attempt to construct $\bar{h}_{lm}^{(i)}(t,r)$ at (or near) the particle through a Fourier sum over modes $R_{lmn}^{(i)}(r)$ will be hampered by the Gibbs phenomenon.

Barack, Ori and Sago \cite{Barack-Ori-Sago} proposed a technique for overcoming this difficulty, named {\it method of extended homogeneous solutions} (MEHS). In Ref.\ \cite{Barack-Ori-Sago} they formulated the method for the scalar-field equation, and worked through a numerical example in which the monopole contribution to the scalar field was calculated for a particle in an eccentric orbit about a Schwarzschild black hole. Later, Hopper and Evans \cite{Hopper-Evans} applied MEHS to the problem of computing the metric perturbation in the Regge-Wheeler gauge. Their treatment was based on the Regge--Wheeler--Zerilli master-function formulation, which reduces the perturbation equations to two separate scalar-like ODEs, one for each parity. Most recently \cite{Hopper-Evans:odd_sector} Hopper and Evans went on to develop a variant of MEHS (dubbed ``method of extended {\it particular} solutions'') that generalizes MEHS to ODEs with non-compact sources. They used this method to tackle the (odd-parity sector of the) gauge transformation equations from the Regge-Wheeler gauge to the Lorenz gauge. MEHS was also employed in Ref.~\cite{Warburton-Barack:eccentric} to compute the scalar-field self-force on a particle moving in the equatorial plane of a Kerr black hole.

In this section we generalize MEHS to the case of multiple coupled fields, relevant to our Lorenz-gauge treatment of the metric perturbation equations. This extension has already been carried out for the monopole ($l=0$) and the dipole ($l=1$) modes by Golbourn \cite{Golbourn-thesis} and implemented for these modes by Barack and Sago \cite{Barack-Sago-eccentric}. Here we present it for a generic $lmn$-mode. We will prescribe, without proof, the construction of the fields $\bar{h}_{lm}^{(i)}(t,r)$ and their derivatives at the particle via MEHS. A proof would closely follow the argument given in \cite{Barack-Ori-Sago}.

For a given $lmn$ mode, the field equations (\ref{eq:metric_pert_FD}) are a set of $k$ coupled second-order ODEs, where $k=1$--$5$ depending on the mode in question (cf.\ Table \ref{table:hierarchical_structure}). We assume that there exists a set of $k$ linearly independent homogeneous solutions $R^{(i)+}_{j}$ ($j=1,\ldots, k$) that satisfy the physical boundary conditions at $r\to\infty$, and another set of $k$ homogeneous solutions $R^{(i)-}_{j}$, linearly independent of each other and of $R^{(i)+}_{j}$, that satisfy the physical boundary conditions at $r\to 2M$. (Here, and in the following discussion, we omit the label $lmn$ for brevity.) The combined set $R^{(i)\pm}_{ j}$ form a complete (2$k$-dimensional) basis of linearly independent solutions to the homogeneous part of Eq.\ (\ref{eq:metric_pert_FD}). That the two sets $R^{(i)+}_{j}$ and $R^{(i)-}_{j}$ exist for each $lmn$ mode of the metric perturbation equations will be confirmed in our analysis by direct construction (the mode $l=m=n=0$ is somewhat exceptional; it will be discussed separately in Appendix \ref{App:monopole}).

Given the basis solutions $R^{(i)\pm}_{ j}(r)$ and their radial derivatives $\partial_{r_*} R^{(i)\pm}_{ j}(r)$, let us construct the $2k\times2k$ matrix 
\begin{eqnarray}\label{eq:Phi_matrix}
	\Phi(r) = \left(\begin{array}{c | c}-R^{(i)-}_{j} & R^{(i)+}_{j}\\ \hline -\partial_{r_*} R^{(i)-}_{j} & \partial_{r_*} R^{(i)+}_{j} \end{array}\right)\c
\end{eqnarray}
where in each $k\times k$ quadrant the rows run over $i$ and the columns run over $j$.
We then define the {\em extended homogeneous solutions} (EHS) as
\begin{eqnarray}\label{eq:radial_EHS_fields}
	\tilde{R}^{(i)}_\pm (r) = \sum_{j=1}^k C^\pm_j R^{(i)\pm}_{j}(r)\c
\end{eqnarray}
where the weighting coefficients $C_j^\pm$ are computed via the matrix equation
\begin{eqnarray}\label{eq:weighting_coeffs}
	\begin{pmatrix} C^-_j \\ C^+_j \end{pmatrix} = \int^{r_\text{max}}_{r_\text{min}} \Phi^{-1}(r) \begin{pmatrix} {\bf 0} \\  J^{(j)}(r) \end{pmatrix} \, f(r)^{-1} dr\p
\end{eqnarray}
Here, the source vector of length $2k$ is formed of $k$ zeroes followed by the $k$ FD sources $J^{(j)}(r)$  of Eq.\ (\ref{eq:metric_pert_FD}) (given explicitly in Appendix \ref{apdx:field_eqs_and_sources}). The factor of $f^{-1}$ comes from the change of integration variable: $dr_\ast = f^{-1}(r) dr $. Note that the EHS functions $\tilde{R}^{(i)}_- (r)$ constitute a solution to the inhomogeneous equations (\ref{eq:metric_pert_FD}) only in the vacuum region $r<r_{\rm min}$; and, similarly, $\tilde{R}^{(i)}_+ (r)$ constitute a solution to the inhomogeneous equations only in the vacuum region $r>r_{\rm max}$. Both EHS functions fail to solve the inhomogeneous equations inside the libration region $r_{\rm min}<r<r_{\rm max}$.

In the final step we define the {\em time-domain} EHS 
fields $\tilde{h}^{(i)lm}_\pm(t,r)$ via the standard Fourier summation
\begin{eqnarray}\label{eq:EHS_TD}
	\tilde{h}^{(i)lm}_\pm(t,r) = \sum_n\tilde{R}^{(i)lmn}_\pm(r) e^{-i\omega t}\p
\end{eqnarray}
The main result of MEHS is that the true time-domain solution, satisfying the physical boundary conditions, is given simply by
\begin{eqnarray} \label{eq:TD}
	\bar{h}^{(i)lm}(t,r) = \left\{\begin{array}{c} \tilde{h}^{(i)lm}_+(t,r), \qquad r > r_p(t), \\ \tilde{h}^{(i)lm}_-(t,r), \qquad r < r_p(t). \end{array} \right.
\end{eqnarray}
Note that, even though each individual EHS function $\tilde{R}^{(i)}_\pm (r)$ fails to be a solution inside libration region, their Fourier mode sums recover the correct time-domain solutions on the corresponding sides of the worldline. The explanation for this result is a straightforward generalization of the argument given in \cite{Barack-Ori-Sago}. 

The main advantage of MEHS is in the fact that the values $\bar{h}^{(i)lm}(t,r_p)$ and $\partial_r\bar{h}^{(i)lm}(t,r_p^{\pm})$, needed as input for the GSF calculation, are obtained via a sum over {\em smooth}, homogeneous Fourier modes. As a result, one encounters no complications related to the Gibbs phenomenon.

As a practical note, we mention that the integral in Eq.\ (\ref{eq:weighting_coeffs}) becomes subtle near the ends of the integration domain, where $J^{(j)}\propto 1/u^r$ and the integrand diverges. We solve this by transforming to $\chi$ as an integration variable:
\begin{eqnarray}\label{eq:weighting_coeffs2}
	\begin{pmatrix} C^-_j \\ C^+_j \end{pmatrix} = \int^{\pi}_{0} \Phi^{-1}(\chi) \begin{pmatrix} 0 \\ \hat J^{(j)}(\chi) \end{pmatrix}\frac{d\tau}{dt} \frac{dt}{d\chi}\, \left(f(r_p(\chi))\right)^{-1} d\chi\c
\end{eqnarray}
where $\hat J^{(j)}\equiv J^{(j)}u^r$ is bounded anywhere in the integration domain. In this expression, $dt/d\chi$ is given in terms of $\chi$ in Eq.\ (\ref{eq:dt_dchi}), and $d\tau/dt=f(r_p(\chi))/{\cal E}$, where $r_p(\chi)$ is given in Eq.\ (\ref{eq:r_p}); both factors are bounded anywhere in the integration domain.

In Appendix \ref{App:monopole} we demonstrate the application of the above method for a particular mode of the perturbation: the mode $l=m=n=0$, i.e.\ the static piece of the monopole perturbation, where the entire construction can be carried out analytically. We choose to discuss this particular mode also because its treatment involves certain subtleties that need to be explained.

\section{The gravitational self-force}\label{sec:GSF}

Detweiler and Whiting \cite{Detweiler-Whiting} showed that, in a local neighborhood of the particle, the retarded Lorenz-gauge metric perturbation $\bar{h}_{\alpha\beta}$ can be split in the form 
\begin{equation}
	\bar{h}_{\alpha\beta} = \bar{h}^R_{\alpha\beta} + \bar{h}^S_{\alpha\beta},
\end{equation}
where the ``singular'' piece $\bar{h}^S_{\alpha\beta}$ is a certain solution to the sourced Lorenz-gauge field equation \eqref{eq:linearized_einstein}, such that (1) the ``regular'' field $\bar{h}^R_{\alpha\beta}$ is a smooth vacuum solution, and (2) the force exerted by $\bar{h}^R_{\alpha\beta}$ on the particle is the physical GSF, as derived by others via rigorous methods \cite{Mino-Sasaki-Tanaka, Quinn-Wald, Poisson-review, Gralla-Wald, Pound}.   
Explicitly, given the field $\bar{h}^R_{\alpha\beta}(x)$ in the particle's neighbourhood, the GSF is given by
\begin{equation}\label{eq:F_self_all_forces}
	F^\alpha_\self =  \mu k^{\alpha\beta\gamma\delta}\nabla_\delta\bar{h}^R_{\beta\gamma}
\end{equation}
(evaluated on the particle), where
\begin{equation}\label{eq:k}
	k^{\alpha\beta\gamma\delta} = \half g^{\alpha\delta}u^\beta u^\gamma - g^{\alpha\beta}u^\gamma u^\delta - \half u^\alpha u^\beta u^\gamma u^\delta + \frac{1}{4}u^\alpha g^{\beta\gamma}u^\delta + \frac{1}{4}g^{\alpha\delta}g^{\beta\gamma}	\p
\end{equation}
The operator $k^{\alpha\beta\gamma\delta}\nabla_\delta$ arises simply from taking the linear-in-$h$ piece of the connection coefficients, and projecting orthogonally to the particle's worldline \cite{Barack:GSF_by_mode_sum}; it is the same operator as the one describing the ``gravitational force'' due to a smooth external perturbation (e.g., an incident gravitational wave) if one projects the motion onto the background spacetime.

Obtaining the field $\bar{h}^R_{\alpha\beta}$ near the particle is the main computational task of any  GSF calculation. A variety of practical approaches have been proposed (see Ref.~\cite{Barack-review} for a review). In this work we make use of the mode-sum method, first proposed in Ref.\ \cite{mode-sum-orig}.

\subsection{GSF via mode-sum regularization}

The mode-sum approach is a practical reformulation of the standard, rigorous GSF formula. Roughly speaking, in this approach the subtraction of the singular field from the full (retarded) field is carried out mode by mode in a multipolar expansion around the large black hole, with the advantage that each modal contribution to the GSF is bounded at the particle. At the operational heart of the method is the {\it mode-sum formula}, 
\begin{eqnarray}\label{eq:ret-mode-sum}
	F^{\alpha}_\text{self} = \sum_{l=0}^\infty \left( F^{\alpha l}_{\pm} - A^\alpha_{\pm} L - B^\alpha - C^\alpha/L \right) \: - D^\alpha \c
\end{eqnarray}
which we now explain.
The quantities $F^{\alpha l}_{\pm}$ are the multipole modes (evaluated at the particle) of the ``full force'' field 
\begin{equation}
F^{\alpha}_{\rm full}\equiv \mu \bar k^{\alpha\beta\gamma\delta}(x)\nabla_\delta\bar{h}_{\beta\gamma}(x),
\end{equation}
 where the field $\bar k^{\alpha\beta\gamma\delta}(x)$ is defined through a certain smooth extension of the quantity $k^{\alpha\beta\gamma\delta}$ off the worldline. By a ``multipole mode'' we mean the quantity obtained by expanding each vectorial component of $F^{\alpha}_{\rm full}$ in {\em scalar} spherical harmonics, and then adding up all $m$-mode contributions for a given $l$. The resulting $l$ modes $F^{\alpha l}_{\pm}$ (which depend on the $k$-extension chosen) turn out to be bounded even at the particle limit, but in general the two radial limits $r\to r_p^{\pm}$ yield two different values---hence the subscript $\pm$. The other terms in the sum in Eq.\ (\ref{eq:ret-mode-sum}) are regularization counter-terms, with 
\begin{equation}
L\equiv l+1/2 \p
\end{equation}
The coefficients $A_\alpha, B_\alpha, C_\alpha, D_\alpha $ are $l$-independent, analytically given {\it regularization parameters}, the values of which are known for generic orbits in Schwarzschild \cite{mode-sum-orig} and Kerr \cite{Barack-Ori} geometries. In the Schwarzschild case they read
\begin{align}
	A^t_\pm 	&= \mp \frac{\mu^2 u^r}{r_p^2 f_p U}\c \qquad A^r_\pm = \mp \frac{\mu^2 \en}{r_p^2 U}\c \qquad A^\varphi_\pm = 0\c	 \label{eq:A^t}	\\
	B^t				&= \frac{\mu^2\en u^r}{\pi r_p^2 f_p U^{3/2}}\left[-K(w) + 2(1-U)E(w)\right],			\\
	B^r				&= -\frac{\mu^2}{\pi r_p^2 U^{3/2}}\left[\left(\en^2+f_p U\right)K(w)-\left[2\en^2(1-U)-f_pU(1-2U)\right]E(w)\right],	\\
	B^\varphi	&= \frac{\mu^2 u^r}{\pi \ang r_p^2 \sqrt{U}}\left[K(w) - \left(1-2\frac{\ang^2}{r_p^2}\right)E(w)\right]	,	\label{eq:B^phi}\\
	C^\alpha	&= 0 = D^\alpha ,
\end{align}
where $K(w)$ and $E(w) \equiv \int^{\pi/2}_0 (1-w\sin^2 x)^{1/2} dx$ are the complete elliptic integrals of the first and second kind respectively, $f_p\equiv f(r_p)$, and
\begin{equation}
	w = \frac{\ang^2}{r_p^2+ \ang^2}\c\qquad U = 1+\frac{\ang^2}{r_p^2}.
\end{equation}
The values of the regularization parameters depend on the $k$-extension chosen, and it is essential that the extension in which the parameters are calculated correspond to that of the (numerically computed)  modes $F^{\alpha l}_{\pm}$. The parameter values we give above correspond to the extension applied in Ref.\ \cite{Barack-Sago-eccentric}, which is defined as follows: For a given particle point $x_p$, take the field  $\bar k^{\alpha\beta\gamma\delta}(x)$ to be given by Eq.~\eqref{eq:k}, with the metric $g^{\alpha\beta}$ taking its value at the field point $x$, and with $u^{\alpha}\equiv u^\alpha(x_p)$ for all $x$ (in Schwarzschild coordinates). Below we will note the practical advantage of this particular $k$-extension. 

The mode sum in Eq.\ \eqref{eq:ret-mode-sum} converges slowly, as $\sim 1/l$. In practice, this means that one has to compute many $l$-modes, which can be computationally demanding. 
It is possible to improve the convergence rate of the mode sum by including higher-order counter terms. It is convenient to choose such terms to have the form
\begin{equation}\label{eq:D_alpha2n}
	\frac{D_{\alpha,2}}{(2l-1)(2l+3)} + \frac{D_{\alpha,4}}{(2l-3)(2l-1)(2l+3)(2l+5)} + \dots = \sum_{N=1}^{N_{\max}} 4^{-N} D_{\alpha,2N}\left[\prod_{k=1}^N(L^2-k^2)\right]^{-1}
\end{equation}
in which the sum $\sum_{l=0}^{\infty}$ of each $N$-term vanishes; for instance, $\sum_{l=0}^\infty[(2l-1)(2l+3)]^{-1}=0$. The coefficients $D_{\alpha,2N}$ are ``high-order regularization parameters''. The $N$th term is proportional to $l^{-2N}$ at large $l$, and no odd powers of $1/L$ occur \cite{Detweiler-Messaritaki-Whiting}. Hence, with the inclusion of each extra parameter, the convergence rate of the mode-sum improves by a factor of $l^{-2}$.
Detweiler \etal~\cite{Detweiler-Messaritaki-Whiting} derived an analytic expression for $D_2$ for the scalar-field self-force on a particle in a circular orbit around a Schwarzschild black hole. \alert{Very recently Heffernan \etal~\cite{Heffernan-Ottewill-Wardell:Schwarz,Heffernan-Ottewill-Wardell:Kerr} were able to derive $D_2$, $D_4$ and $D_6$ for the scalar-field self-force and $D_2$, $D_4$ for the electromagnetic and gravitational self-forces, in all cases for generic orbits in Schwarzschild or Kerr geometry}. We will make use of their results in our computation.

The mode-sum formula \eqref{eq:ret-mode-sum} requires as input the {\it scalar} spherical harmonic modes of the components $F^{\alpha}_{\rm full}$. These are to be constructed from numerically-computed {\it tensor} harmonic modes of the metric perturbation. This involves a projection of the functions $F^{\alpha}_{\rm full}(t,r,\theta,\varphi)$ (for each Schwarzschild component $\alpha$ treated as a scalar field) onto a basis  of scalar harmonics. The outcome will, of course, depend on the off-worldline extension chosen for $k^{\alpha\beta\gamma\delta}$. Generically, each spherical-harmonic $l$ mode of $F^{\alpha}_{\rm full}$ will couple to infinitely many tensor-harmonic modes of the input perturbation. To minimize the level of coupling requires a judicious choice of the $k$-extension.
With the choice made above, following Ref.\ \cite{Barack-Sago-eccentric}, only a finite coupling occurs: in general, each scalar-harmonic $l$-mode $F^{\alpha l}_{\pm}$ has contributions from only seven tensor-harmonic $l'$-modes, i.e. $l-3\le l' \le l+3$ (and there is no coupling between different $m$-modes).
The resulting formula for $F^{\alpha l}_{\pm}$ has the form \cite{Barack-Sago-eccentric}
\begin{align}\label{eq:tensor_scalar_coupling}
	F^{\alpha l}_{\pm} = \frac{\mu^2}{r_p^2}\sum_{m=-l}^l&\left\{\mathcal{F}^{\alpha l-3,m}_{(-3)} + \mathcal{F}^{\alpha l-2,m}_{(-2)} + \mathcal{F}^{\alpha l-1,m}_{(-1)} + \mathcal{F}^{\alpha l,m}_{(0)} \right.\\
		&+ \left.\mathcal{F}^{\alpha l+1,m}_{(+1)} +\mathcal{F}^{\alpha l+2,m}_{(+2)} +\mathcal{F}^{\alpha l+3,m}_{(+3)} \right\}Y^{lm}(\theta_p,\varphi_p)\c \nonumber
\end{align}
where each quantity $\mathcal{F}^{\alpha l+k,m}_{(n)}$ (with $ k =-3, \ldots 3$) is constructed from the tensor-harmonic mode $\bar h^{l+k,m}_{\alpha\beta}(t,r)$ of the metric perturbation, and its first derivatives. Explicit expressions for the ${\cal F}s$ are given in Appendix C of Ref.\ \cite{Barack-Sago-eccentric}; 
we do not repeat them here as they are rather lengthy.

\subsection{Conservative and dissipative components of the GSF}\label{sec:cons_diss}

When analyzing the different physical effects of the GSF it is physically useful to consider its conservative and dissipative effects separately \cite{Barack-review,Hinderer-Flanagan}. Splitting the GSF into its conservative and dissipative components, $F_\alpha^\text{cons}$ and $F_\alpha^\text{diss}$, is also practically beneficial, as the two pieces admit $l$-mode sums with different convergence properties, which are better dealt with indepentently. In the case of bound orbits around a Schwarzschild black hole, one may readily extract the dissipative and conservative pieces of the GSF taking advantage of the orbital symmetries. As discussed in Ref.~\cite{Barack-review}, assuming (without loss of generality) that $\tau=0$ corresponds to a periastron passage, we can write
\begin{eqnarray} \label{eq:diss-cons}
	F_\alpha^\text{cons}(\tau) = \frac{1}{2} [ F_\alpha^\text{ret}(\tau) + \epsilon_{(\alpha)} F_\alpha^\text{ret}(-\tau)], 	\qquad		F_\alpha^\text{diss}(\tau) = \frac{1}{2} [ F_\alpha^\text{ret}(\tau) - \epsilon_{(\alpha)} F_\alpha^\text{ret}(-\tau)] 
\end{eqnarray}
(no summation over $\alpha$), where $\epsilon_{(t,\varphi)}=-1$ and $\epsilon_{(r)}=+1$. In our analysis we assume the orbit is equatorial, so $F_\theta^\text{cons}= 0$ as well as $F_\theta^\text{diss}= 0$ from symmetry.

One of the advantages of splitting the GSF in this manner is that the dissipative piece of the GSF does not require regularization. The conservative piece, on the other hand, is regularized with the same regularization parameters as the complete GSF. Explicitly, the mode-sum formulas for the conservative and dissipative pieces are given by \cite{Barack-review}
\begin{eqnarray}
	F_\alpha^{\text{cons}} &=& \sum_{l=0}^\infty \left[ F_{\alpha l \pm}^{\text{full(cons)}} - A_\alpha^\pm L - B_\alpha \right]		\label{eq:cons-regularization}	\c\\
	F_\alpha^{\text{diss}} &=& \sum_{l=0}^\infty F_{\alpha l \pm}^{\text{full(diss)}}		\c											\label{eq:diss-regularization}
\end{eqnarray}
where $F_{\alpha l \pm}^{\text{full(cons)}}$ and $F_{\alpha l \pm}^{\text{full(diss)}}$ are constructed using formulas analogous to Eqs.~(\ref{eq:diss-cons}), and we have used $C_{\alpha}=0$. The same higher-order regularization parameters used to improve the convergence of the complete GSF can also be used to improve the convergence of $F_\alpha^{\text{cons}}$.

\section{Numerical implementation}\label{sec:numerical_implemenation}

\subsection{Numerical boundary conditions} \label{sec:numerical_BC}

Our primary numerical task is to solve the radial equation \eqref{eq:metric_pert_FD} subject to the boundary conditions given by Eq.\ (\ref{eq:physical_BC_in_FD}). In our numerical implementation we use the radial coordinate $r_*$, in terms of which the physical boundaries are located at $\pm\infty$, so integrating strictly from the physical boundaries is not possible. Instead, we place approximate boundary conditions at the edges of our numerical domain, which runs (for a given $\omega$ mode) from $r_\ast=r_*^\text{in} \ll -|\omega|^{-1}$ to $r_\ast=r_*^\text{out} \gg \omega^{-1}$ (for static modes, $\omega=0$, these conditions are replaced with $r_*^\text{in} \ll -M$ and $r_*^\text{out} \gg M$). How we choose $r_*^\text{in}$ and $r_*^\text{out}$ in practice will be discussed in the next subsection.
Approximate boundary conditions for the homogeneous fields $R^{(i)}_{lm\omega}$ were developed in Paper I, and we adopt them here. We will review here the form of these boundary conditions, and refer the reader to Paper I for further details.

In constructing the numerical boundary conditions it is assumed, a priori, that the radial fields admit an asymptotic expansion in $1/r$ at $r\rightarrow\infty$ and an asymptotic expansion in $r-2M$ at $r\rightarrow 2M$. Combined with the leading-order behavior of the physical perturbation given in Eq.~(\ref{eq:physical_BC_in_FD}), this leads to the ansatz 
\begin{align}
	R^{(i)}_-(r_\text{in})  &= \exp(-i\omega r_*^\text{in}) \sum^\infty_{j=0} b^{(i)}_j(r_\text{in}-2M)^j		\c	\label{eq:inner_fields_ansatz} \\
	R^{(i)}_+(r_\text{out}) &= \exp(+i\omega r_*^\text{out}) \sum^\infty_{j=0} \frac{a^{(i)}_j}{r_\text{out}^j}	\c	\label{eq:outer_fields_ansatz}
\end{align}
where $r_\text{in}\equiv r(r_*^\text{in})$ and $r_\text{out}\equiv r(r_*^\text{out})$, and we hereafter suppress the indices $lm\omega$ for brevity. The above ansatz turns out to be appropriate for all $lm\omega$ modes, with the exception of the even-parity static modes (i.e, $l+m$=even with $\omega=0$) of $R^{(i)}_+$, to be discussed separately below. By substituting the above ansatz into the field equations \eqref{eq:metric_pert_FD}, one obtains recursion relations between the $a^{(i)}_j$'s and (separately) between the $b^{(i)}_j$'s. These relations are given in Paper I. For each $lm\omega$, there are precisely $k$ freely specifiable parameters $a^{(i)}_j$ and $k$ more freely specifiable parameters $b^{(i)}_j$, where $2k$ is the dimension of the space of homogeneous solutions for that mode (cf. Table \ref{table:hierarchical_structure}). If we arrange these freely specifiable parameters to form vectors $\vec a=\{a_1,a_2,\ldots,a_k\}$ and $\vec b=\{b_1,b_2,\ldots,b_k\}$, then by choosing $k$ linearly independent vectors $\vec a$ ($\vec b$) we obtain a basis of $k$ linearly independent homogeneous solutions $R^{(i)}_+$ ($R^{(i)}_-$).

For even-parity static modes, the ansatz in Eq.\ \eqref{eq:outer_fields_ansatz} does not produce the necessary number of freely specifiable parameters $a^{(i)}_j$ ($k=3$ in this case---recall Table \ref{table:hierarchical_structure}).
For these modes, one instead uses 
\begin{equation}
	R^{(i)}_+(r_\text{out}) = \sum^{\infty}_{j=l} \frac{a^{(i)}_j + \bar{a}^{(i)}_j \ln (r_\text{out}/M)}{r_\text{out}^j}	\label{BC_static_even}	\c
\end{equation}
which gives three freely specifiable parameters as required (these can be taken to be $\vec a=\{a_l^{(3)},a_l^{(5)},a_{l+2}^{(5)}\}$, which then determines all the other $a^{(i)}_j$ and $\bar{a}^{(i)}_j$ coefficients). See paper I for details.

\subsection{Computational algorithm}

We now outline the necessary steps in computing the Lorenz-gauge GSF for a particle on an eccentric orbit, in our FD approach. The algorithm is similar to that used in Ref.~\cite{Warburton-Barack:eccentric} for the scalar-field self-force. 

\begin{enumerate}
\item{\it Orbital parameters.} For given orbital eccentricity, $e$, and semi-latus rectum, $p$, we calculate relevant orbital parameters ($\en, \ang, \Omega_r, \Omega_\varphi,T_r\ \text{etc.})$ using the formulas given in Sec.~\ref{sec:Schwarzschild_geodesics}.

\item{\it Boundary conditions.}
For a given $lmn$ mode we derive boundary conditions using Eqs.\ (\ref{eq:inner_fields_ansatz}) and (\ref{eq:outer_fields_ansatz}) [or, for static modes of even parity, Eq.\ (\ref{BC_static_even})].  
On the outer boundary we derive $k$ sets of values $\left.\{R^{(i)}_{+},\partial_{r_*} R^{(i)}_{+}\}\right|_{r^*_{\rm out}}$, corresponding to $k$ linearly independent choices of $\vec a$; the choice $\vec a_1=(1,0,\ldots,0)$, $\vec a_2=(0,1,\ldots,0),\ldots$ is convenient. We similarly obtain $k$ sets of values $\left.\{R^{(i)}_{-},\partial_{r_*} R^{(i)}_{-}\}\right|_{r^*_{\rm in}}$ corresponding to $k$ linearly independent choices of $\vec b$. On each boundary there are two control parameters: the value of $r_*^{\rm in/out}$ and the truncation index $j=j_{\rm max}$. We choose these values so as to achieve a relative error $<10^{-14}$ in the partial sums for $R^{(i)}_\pm(r^*_{\rm in/out})$. Since evaluating the boundary conditions is substantially cheaper than integrating the field equations, it is advantageous to  place the boundaries as close to the particle as possible, at the modest cost of increasing $j_{\rm max}$.
Through experimentation we found that setting $r^\text{out}_* = 10/\omega$ and $r_*^\text{in}=-50M$ worked well in most cases we considered. With these values we usually needed to truncate the series at $j_{\max} \lesssim 15 $ for the outer boundary and at $j_{\max} \lesssim 5$ for the inner boundary.

\item{\it Homogeneous fields.} 
For our given $lmn$, and for each one of the $k$ sets of boundary values at $r_{\rm out}$, we integrate the homogeneous part of the coupled field equations (\ref{eq:metric_pert_FD}) from $r_*=r^*_{\rm out}$ inward to $r_*^\text{min}$. Similarly, for each one of the $k$ sets of boundary values at $r_{\rm in}$, we integrate the homogeneous ODEs from $r_*=r^*_{\rm in}$ outward to $r_*^\text{max}$. We use the Runge-Kutta Prince-Dormand integration routine (rk8pd from \cite{GSL}). (For $l=m=n=0$ the homogeneous solutions are obtained analytically, as prescribed in Appendix B.) This yields a $2k$-dimensional basis of homogeneous solutions $R^{(i)}_{\pm}(r)$. For our purpose it suffices to record the values of these fields (and the values of their $r_*$ derivatives) in the libration interval, $r_*^\text{min}\leq r_*\leq r_*^\text{max}$. In our implementation we record these values at 5000 radii in this interval, equally spaced in $r_*$.

\item{\it Extended homogeneous solutions.} For our given $lm\omega$, we construct the EHS $\tilde R^{(i)}_{\pm}(r)$ using Eqs.~\eqref{eq:radial_EHS_fields} and \eqref{eq:weighting_coeffs2}.
We find it important that the integration in Eq.\ \eqref{eq:weighting_coeffs} is performed to a high accuracy, a somewhat challenging task due to the oscillatory nature of the integrand. We achieve sufficiently high accuracy by coupling a standard adaptive integrator routine from the GNU Scientific Library (GSL) \cite{GSL} to the ODE solver. When the integrator requests the value of the integrand at a particular value of $r_*$, the ODE solver is loaded with information from the nearest of the 5000 points stored in the previous step, and then it integrates the homogeneous fields from that point up to the requested $r_*$ value. 
(We find that simply interpolating the data stored in the previous step does not produce sufficiently accurate results.) For modes with very small values of $|\omega|$, the matrix $\Phi$ becomes nearly singular and thus difficult to invert. Such modes are dealt with separately as discussed in Sec.\ \ref{sec:nearly_static_modes} below.  
Finally, following the hierarchical scheme illustrated in Table \ref{table:hierarchical_structure}, we use the gauge equations \eqref{eq:gauge1} to \eqref{eq:gauge4} to construct any remaining EHS fields that have not been computed via integration of the field equations (e.g., for $l\geq 2$ with $\omega=0$, we use the gauge conditions to construct $\tilde R^{(6,7)}_{\pm}$ out of $\tilde R^{(1,3,5)}_{\pm}$).

\item{\it Sum over $\omega$ modes.} Steps 2 through 4 are repeated for sufficiently many $n$ modes, and then the time-domain EHS fields $\tilde{h}^{(i)lm}_\pm(t,r)$ are constructed via Eq.\ \eqref{eq:EHS_TD}. The $n$-mode sum converges exponentially fast, since each of the modes is a homogenous (smooth) solution---this is the main advantage of the EHS method. To decide where to truncate the sum over $n$, we take advantage of the fact that the EHS fields $\tilde R_{\pm}^{(i)}$ are {\it not} solutions of the inhomogeneous FD field equations in the libration region, while the time-domain field $\tilde{h}^{(i)lm}(t,r)$ defined in Eq.\ (\ref{eq:TD}) {\it is} everywhere a solution of the corresponding inhomogeneous time-domain equations. In particular, the functions $\tilde R_{\pm}^{(i)}$ generally fail to match continuously on the particle's worldline, while their time-domain counterparts $\tilde{h}^{(i)lm}_\pm$ do match there. Similarly, the jump in the derivatives of the $\pm$ fields at the worldline is consistent with the distributional source of the field equations only upon summation over $n$. Therefore, for a given partial sum $\sum_{n=0}^{n_{\max}}$, the residuals $[\tilde R^{(i)}]_{r_p}\equiv \tilde R_{+}^{(i)}(r_p(t))-\tilde R_{-}^{(i)}(r_p(t))$ or $[\tilde R^{(i)}_{,r}]_{r_p}\equiv \tilde R_{+,r}^{(i)}(r_p(t))-\tilde R_{-,r}^{(i)}(r_p(t))$ (as compared to the expected jump in $\tilde{h}^{(i)lm}$) can serve as measures of the truncation error. We may control this error by thresholding on these residuals. In our implementation we set a threshold of $10^{-12}$ on the relative difference between $[\tilde R^{(i)}_{,r}]_{r_p(t)}$ and the expected jump, maximized in $t$ over an entire radial period. Once the threshold is reached, we record $\tilde{h}^{(i)lm}(t,r_p)$ and the two sided values of $\tilde{h}_{,r}^{(i)lm}(t,r_p)$ along the worldline. These will be used as input for the GSF calculation in the last step.

The necessary number of modes $n_{\max}$ (for a given threshold) is a function of $l,m$ and of the orbital parameters; it depends particularly strongly on the eccentricity $e$. In Sec.\ \ref{sec:results} below we will provide some indicative information about the number of modes required in practice.

\item{\it Sum over $lm$-modes, and the GSF.} 
Step 5 is repeated for sufficiently many $l,m$ modes (note that there is no need to compute the $lm$ modes in any particular order---in Sec.\ \ref{sec:parallelization} we discuss how this element of the computation can be parallelized ). Using Eq.\ \eqref{eq:tensor_scalar_coupling} and the expressions from Appendix C of Ref.\ \cite{Barack-Sago-eccentric} we compute $F^{l(\full)}_\alpha$, the $l$-mode contribution to the full force. Recall that, owing to the coupling between scalar and tensor modes, if we wish to calculate a given $l_\max$ number of scalar $l$-modes we must calculate $l_{\max}+3$ tensor modes [see Eq.\ \eqref{eq:tensor_scalar_coupling}]. The different convergence properties of the conservative and disspative components of the GSF make it beneficial to split each mode $F^{l(\full)}_\alpha$ into these two components using (the $l$-mode version of) Eq.\ \eqref{eq:diss-cons}. The two pieces $F_\alpha^{\rm cons}$ and $F_\alpha^{\rm diss}$ are then obtained using the mode-sum formulas (\ref{eq:cons-regularization}) and (\ref{eq:diss-regularization}), with the regularization parameters given in Eqs.\ (\ref{eq:A^t})--(\ref{eq:B^phi}). To improve the convergence of the mode-sum for the conservative piece, we also use the high-order parameters $D_2$ and $D_4$ from Heffernan \etal\ \cite{Heffernan-Ottewill-Wardell:Schwarz}.

In the case of $F_\alpha^{\rm diss}$ the mode-sum converges exponentially fast in $l$, and we 
find that $l_\max=15$ typically suffices to obtain the dissipative GSF to within our target relative accuracy of $10^{-6}$. In the case of $F_\alpha^{\rm cons}$  the convergence is slower (see below), and it is in some cases necessary to estimate the contribution from the truncated large-$l$ tail. Our method for doing so is detailed in the next subsection.  

\end{enumerate}

\subsection{Contribution from truncated large-$l$ tail}
In practice, for moderate eccentricities, it is computationally prohibitive to numerically calculate modes with $ l $ much larger than $\sim 20 $. We estimate the contribution from the large-$l$ modes by a fitting scheme. Let us focus on the conservative component of the GSF, for which the issue becomes a problem.
We write this component as a sum of two pieces, a numerically computed partial sum, and a large-$l$ tail:
\begin{eqnarray}
    F_\alpha^{\rm cons} = F_\alpha^{l\leq l_{\text{max}}} + F_\alpha^{l > l_{\text{max}}} \p
\end{eqnarray}
Here
\begin{eqnarray}
    F_\alpha^{l \leq l_{\text{max}}} \equiv \sum^{l_{\text{max}}}_{l=0} F_{\alpha}^{l\text{(reg)}} \qquad \text{and} \qquad F_\alpha^{l>l_{\text{max}}} \equiv \sum^\infty_{l=l_{\text{max}}+1} F_{\alpha}^{l\text{(reg)}} \c
\end{eqnarray}
where the ``regularized'' modes, in our implementation, are given by
\begin{equation}\label{eq:Frl_reg}
	F^{l\text{(reg)}}_\alpha = F^{l(\full)}_\alpha - A_\alpha L - B_\alpha - \sum_{N=1}^{2} 4^{-N} D_{\alpha,2N}\left[\prod_{k=1}^N(L^2-k^2)\right]^{-1} \p		
\end{equation}
Recall $L=l+1/2$, and $\{D_{\alpha,2},D_{\alpha,4}\}$ are the high-order parameters computed analytically in \cite{Heffernan-Ottewill-Wardell:Schwarz}. We expect $F^{l\text{(reg)}}_\alpha$ to fall off as $\sim L^{-6}$ at large $l$, so the truncation tail  $F_\alpha^{l > l_{\text{max}}}$ is expected to fall off like $\sim l_{\max}^{-5}$ for large $l_{\max}$.

The large-$l$ contribution $F_\alpha^{l>l_{\text{max}}}$ can be computed by extrapolating the last few numerically calculated $l$-modes. In our code we use a standard least-squares algorithm from the GSL \cite{GSL} to fit for the coefficients $D_{\alpha,2N}$ (with $N>2$) using our numerical data for $l$-modes with $l_{\max} - 6 \le l \le l_{\max}$, i.e.~the last 7 data points from the set of available numerically computed $l$ modes. When selecting a subset of $l$ modes for the fit, one has, on one hand, to sample from the large-$l$ portion of the data (where the behavior is an approximate $L^{-6}$  drop-off), and on the other hand to be mindful of the fact that large-$l$ modes carry a large relative error. We have experimented with fitting to different subsets of data points, and found that the smallest variance (in the magnitude of the fitted tail) was obtained when fitting the last $5$--$10$ points. We quote here results for a 7-point fit, and use the said variance as a rough measure of the fitting error. 

Given the (numerically fitted) regularization parameters $D_{\alpha,2N}$, we estimate the high-$l$ contribution using the formula
\begin{equation}
	F_\alpha^{l > l_{\text{max}}} = \sum_{N=3}^{N_{\max}} \frac{D_{\alpha,2N} (-4)^{-N}\pi(-1)^{l_{\max}+1}(l_{\max}+1)}{(2N-1)\Gamma(N-l_{\max}-1/2)\Gamma(N+l_{\max}+3/2)},
\end{equation}
where $\Gamma(x)$ is the standard gamma function. With $l_\max=15$ we find it sufficient to take $N_\max=4$ (i.e., fit for $D_{\alpha,6}$ and $D_{\alpha,8}$ only) to reach our target accuracy of $10^{-6}$ relative to the magnitude of $F_\alpha^{\rm cons} $. The residual overlooked by this 2-parameter fit is of the order of $ l_{\max}^{-9}$, which is $\mathcal{O}(10^{-12})$ for $l_{\max} = 20 $. Were $D_{\alpha,6}$ and $D_{\alpha,8}$ to be computed analytically it would no longer be necessary to estimate the large-$l$ tail to reach our desired accuracy goal.


\subsection{Code parallelization}\label{sec:parallelization}

Since each tensorial $lm$-mode of the metric perturbation is calculated independently from others, our computational problem is amenable to parallelization in an obvious manner. 
Our code is written to run on multiple CPUs, either within a single machine or on a cluster, using the Message Passing Interface (MPI). We also make use of dynamic load balancing whereby the root processor forks a thread that keeps record of the $lm$ modes that have already been computed. Once a processor has been assigned an $lm$ mode, it begins computing the $n$ modes in the order $n=0,-1,1,-2,2\dots$ and continues until a convergence criteria is met as discussed above. After a given processor completes an $lm$ mode computation, it contacts the thread on the root processor to request a new mode to work on. Each processor records its calculated contribution to the total GSF, and once all necessary $lm$ modes have been computed the results are combined. This simple parallel algorithm allows us to calculate the GSF rapidly on a cluster of computers---see Fig.~\ref{fig:GSF_run_times}.

\section{Problem of low-frequency modes and its mitigation}\label{sec:nearly_static_modes}

\begin{figure}
	\centering
	\includegraphics[width=100mm]{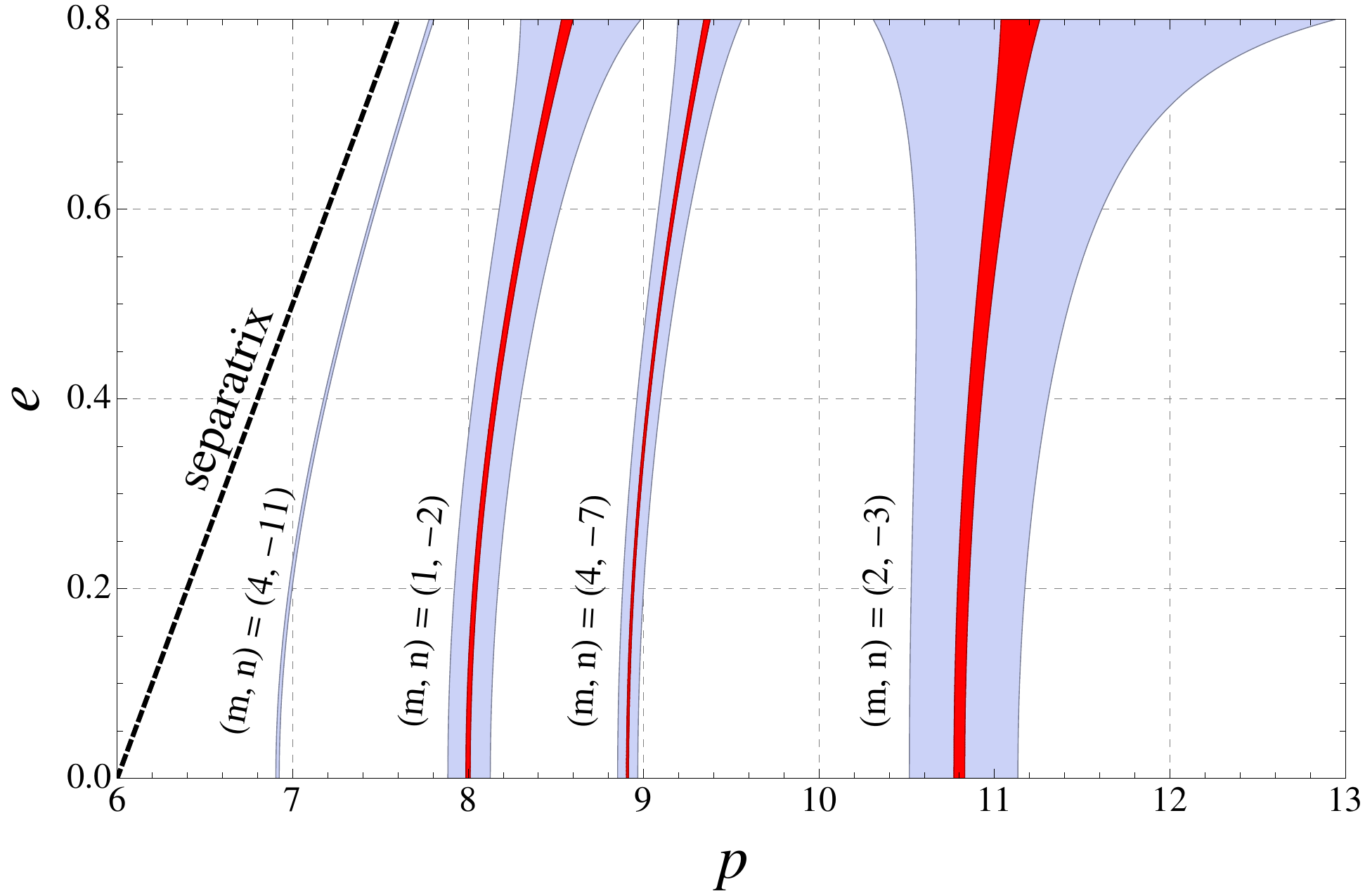}
	\caption{Regions of strong-field parameter space where low-frequency modes are encountered, showing just a sample of low-order $(m,n)$ ``resonances''. The broader (light/violet) bands mark regions where $|\omega|=|m\Omega_\varphi + n \Omega_r|<10^{-3}M^{-1}$, and the narrower (dark/red) bands mark regions where $|\omega|<10^{-4}M^{-1}$. Relevant resonances are those for which the frequency ratio $\Omega_\phi/\Omega_r$ is a rational number $|n/m|$, where $n$ and $m$ are the indices for the modes that need to be calculated numerically (usually small integers). ``Near-resonant'' modes are difficult to deal with numerically, as discussed in the text. Our current code incorporates a method to mitigate this problem, which allows us access to all points in the parameter space except those for which there occur low-order harmonics with $|\omega|\lesssim 10^{-4}M^{-1}$ (narrower bands in the figure).
}
\label{fig:low_omega_regions}
\end{figure}

In our discussion so far we have a assumed that the code does not encounter nonstatic modes of very small frequency, $|\omega|=|m\Omega_\varphi + n \Omega_r|\ll M^{-1}$. In reality, modes of $|\omega|$ small enough to cause numerical difficulties (for reasons described below) will be encountered in a wide portion of the parameter space. Figure \ref{fig:low_omega_regions} shows $\omega={\rm const}$ contours in the $e,p$ space around 4 sample ``resonances'' corresponding to $\Omega_r/\Omega_\varphi=|m/n|=\frac{2}{3}$, $\frac{4}{7}$, $\frac{1}{2}$ and $\frac{4}{11}$. Around each resonance, the wider band marks orbits with $M|\omega|<10^{-3}$, and the narrower band marks orbits with $M|\omega|<10^{-4}$. The bands generally become wider for smaller $|m|+|n|$, larger $p$ and larger $e$. Since $\Omega_\varphi-\Omega_r\sim 3p^{-5/2}$ at large $p$, we find that all orbits with $p\gtrsim 25$ ($\gtrsim 62$) lie within the $M|\omega|<10^{-3}$ ($<10^{-4}$) band for $(m,n)=(\pm 1,\mp 1)$.
We have found through experimentation that our code cannot solve accurately for modes with $M|\omega|\lesssim 10^{-3}$. We have devised a partial remedy to this problem, allowing us to compute modes with frequency as low as $M|\omega|\approx 10^{-4}$. The gain, in terms of accessibility to a larger portion of the parameter space, can be appreciated from Fig.\ \ref{fig:low_omega_regions}. In this section we will diagnose the root causes for the numerical problem at low-frequency, and describe the partial cure we have devised to mitigate it.

The low-frequency problem lies in the numerical task of accurately inverting the matrix $\Phi$ of EHS fields [recall Eq.\ \eqref{eq:weighting_coeffs}]. The matrix becomes ill-conditioned at very small $M|\omega|$, for two independent reasons. First, the dimension of the space of homogeneous solutions that make up $\Phi$ is smaller on the resonance than it is off the resonance: as $\omega\to 0$, some of the functions $\{R^{(i)},\partial_r R^{(i)}\}$ in $\Phi$ become linearly dependent through the gauge conditions (\ref{eq:gauge1})--(\ref{eq:gauge4}). In consequence, the matrix $\Phi$ degenerates in the limit $\omega\to 0$. When $\omega$ is finite but small, $\Phi$ retains its full off-resonance dimensionality, but may become ill-conditioned.

We have found, however, that the singular behavior of $\Phi$ at low $\omega$ is dominated by yet another effect, which can be described as follows. When $|\omega|$ is very small, the transition to a ``wave-zone'' (oscillatory) behavior occurs at very large values of $r$. Between the libration region and the distant wave-zone (where we set our outer boundary $r_{\rm out}$) there is a very wide domain, mostly located in the weak-field regime, in which the fields $R^{(i)}$ decay (essentially) with power laws. It turns out that each $R^{(i)}$ is a combination of power-law terms of different decay rates, so that certain linear combinations of $R^{(i)}$ can exhibit different power-law behavior.  When we integrate our ODEs inward from $r_{\rm out}$ over the large span of space down to the libration region, different power-law tails will develop very different amplitudes. As a result, we will have certain linear combinations of $R^{(i)}$ solutions that are much smaller than the functions $R^{(i)}$ themselves. This manifests itself with large condition numbers and an ill-conditioned matrix $\Phi$. (We will illustrate this behavior below with an example.)

Left untreated, the above problem restricts severely the portion of the parameter space in which our code can work reliably. We find that even-parity modes with $M|\omega|\lesssim 10^{-3}$ cannot be computed even at a moderate accuracy, and, in practice, this leaves a very significant portion of the parameter space out of our reach. We have devised a partial remedy to this situation, addressing the second of the above problems, though not the more fundamental problem of matrix degeneracy. This, however, already allows us to extend the range of computable frequencies down to 
$M|\omega|\approx 10^{-4}$, permitting access to a much larger portion of the parameter space. To improve this further, Ref.\ \cite{Warburton-thesis} describes an idea for a systematic solution to the matrix degeneracy problem, which is based on a perturbative treatment of the ODEs with $\omega$ as a small parameter. The details of this method are yet to be worked out and implemented.

In what follows we first illustrate the low-frequency problem and our suggested partial cure in the particular example of $l=1=m$. We then generalize to modes of arbitrary $l,m$.

\subsection{An example: even-parity dipole mode}\label{sec:basis_rotation_even_dipole}

Consider the mode $l=1=m$, for which one has to solve the (homogeneous part of the) coupled set \eqref{eq:metric_pert_FD} for $\vec{R} \equiv (R^{(1)}, R^{(3)}, R^{(5)}, R^{(6)})^T$. For $\omega\ne 0$ there are two remaining nontrivial functions, $R^{(2)}$ and $R^{(4)}$, which are then obtained algebraically, given $\vec{R}$, using the gauge conditions (\ref{eq:gauge2}) and (\ref{eq:gauge3}), respectively. However, for $\omega=0$ the functions $R^{(2)}$ and $R^{(4)}$ no longer feature in the gauge conditions (\ref{eq:gauge2}) and (\ref{eq:gauge3}), and these equations then form nontrivial linear relations (at each given $r$) between the functions $\{\vec{R},\vec{R}_{,r}\}$. As a result, the corresponding 8-dimensional matrix $\Phi$ degenerates---this is the ``matrix degeneracy'' problem discussed above. Since we have two nontrivial linear relations, we expect the null space of $\Phi$ to be 2-dimensional, and $\det\Phi\propto \omega^2$ at small $\omega$. We have confirmed the expected scaling numerically with our code.

Let us now turn to the second problematic effect of a low frequency, which (left untreated) we have found to be even more restrictive than the matrix degeneracy problem. For $\omega\ll M^{-1}$ we have a large domain $M\ll r\ll 1/\omega$, in which the behavior is both weak-field and non-oscillatory. In this domain, the set of $l=1=m$ ODEs \eqref{eq:metric_pert_FD} takes the (approximate) weak-field form 
\begin{equation}\label{eq:large_r_R}
	\vec{R}''(r) + r^{-2}A_1\vec{R}(r) = 0\c
\end{equation}
where a prime denotes $d/dr$, and
\begin{equation}
A_1= \left( \begin{array}{cccc}
	-4 & 2 & 2 & 2		\\
	2 & -4 & -2 & -2 	\\
	4 & -4 & -6 & -4	\\
	2 & -2 & -2 & -4\end{array}	\right)\p
\end{equation}
The matrix $A_1$ can be diagonalized using $D_1=Q_1^{-1}A_1 Q_1$ with 
\begin{equation}
	Q_1= \left( \begin{array}{cccc}
	1 & 1 & 1 & -1		\\
	1 & 0 & 0 & 1 	\\
	0 & 0 & 1 & 2	\\
	0 & 1 & 0 & 1\end{array}	\right)
\end{equation}
and 
\begin{equation}
D_1=\text{diag}(-2, -2, -2, -12) \p
\end{equation}
In terms of the ``rotated'' basis $\vec{R}_{\rm rot} \equiv Q_1^{-1} \vec{R}$, the set (\ref{eq:large_r_R}) decouples, reading
\begin{equation}\label{eq:vec_F_field_eq}
	\vec{R}''_{\rm rot}(r) + r^{-2}D_1\vec{R}_{\rm rot}(r) = 0\p
\end{equation}
The relevant ```external'' solution is 
\begin{equation}
\vec{R}_{\rm rot}= \left(\frac{c_1}{r},\frac{c_2}{r}, \frac{c_3}{r} , \frac{c_4}{r^3}\right)^T \c
\end{equation}
where $c_i$ are arbitrary amplitudes. We see that, over the domain $M\ll r\ll 1/\omega$, three of the decoupled fields $\vec{R}_{\rm rot}$ behave as $\propto 1/r$ while the fourth has a much steeper tail of $\propto 1/r^3$. If we integrate the  ODEs inward starting in the wave-zone with similar initial amplitudes $c_i$ for all four fields (as we effectively do in practice), then the fourth ($\propto 1/r^3$) field will reach a much higher amplitude in the libration region, making $\Phi$ ill-conditioned. This situation is irrespective of whether we are integrating the original $\vec R$ equations or the rotated $\vec R_{\rm rot}$ equations, since the condition number of $\Phi$ is invariant under rotations in the solution space. We expect the condition number to scale as $\sim r_{\rm out}^{2}$ (and hence roughly as $\sim\omega^{-2}$) due to this effect. 



The above diagnosis suggests an obvious way to mitigate the problem: we need only rescale the initial amplitude of the $\propto r^{-3}$ basis vector, in such a way that once the fields are integrated down to the libration region they all reach similar magnitudes. In terms of the $\vec R_{\rm rot}$ variables, if we choose our initial amplitudes $\{a_0^{(1)},a_0^{(3)},a_0^{(5)},a_0^{(6)}\}$ [recall Eq.\ \eqref{eq:outer_fields_ansatz}] to be $\{1,0,0,0\}$, $\{0,1,0,0\}$, $\{0,0,1,0\}$ and $(M\omega)^2\{0,0,0,1\}$, we find that the solutions $R^{(i)}$ will all reach similar amplitudes in the libration region, as desired. Of course, we need not actually re-express our ODEs in terms of the $\vec R_{\rm rot}$ variables to implement this fix: we may continue to work with the original ODEs, simply adjusting the initial amplitudes to be $(Q_1)_{i1}$, $(Q_1)_{i2}$, $(Q_1)_{i3}$ and $(M\omega)^2(Q_1)_{i4}$, namely $\{1,1,0,0\}$, $\{1,0,0,1\}$, $\{1,0,1,0\}$ and $(M\omega)^2\{-1,1,2,1\}$.
With this choice of initial amplitudes we find that $\Phi$ becomes much better conditioned. Even though this simple technique does not address the fundamental problem of degeneracy, we find that it allows us access to frequencies smaller by at least an order of magnitude. This level of accuracy sufficed for the orbital-evolution application considered in Ref.\ \cite{Warburton_etal}.

\subsection{Modes of general $l,m$}\label{sec:basis_rotation}
The above analysis generalizes to arbitrary $l,m$. For {\it odd-parity} modes with $l\ge1$ we need to solve for the basis $\vec{R}^{\rm odd} \equiv (R^{(9)} , R^{(10)})^T$ (in the case $l=1$ the set of ODEs reduces to a single equation). For $M\ll r\ll 1/\omega$ the set of ODEs reduces to a form similar to that of Eq.\ \eqref{eq:large_r_R}, with the matrix $A_1$ replaced with 
\begin{equation}
	A_l^{\rm odd} = \left( \begin{array}{cc}
		-(\Lambda+4) & 2 \\
		2\Lambda-4 & -(\Lambda-2) 	\end{array}\right)\c
\end{equation}	
where $\Lambda \equiv l(l+1)$. $A_l^{\rm odd}$ is diagonalized using $D_l^{\rm odd}=(Q_l^{\rm odd})^{-1}A_l^{\rm odd} Q_l^{\rm odd}$, with 
\begin{equation}
Q_l^{\rm odd}= \left( \begin{array}{cc}
		\frac{1}{l+2} & -\frac{1}{l-1} \\
	    1 & 1 	\end{array}\right)
\end{equation}
and
\begin{equation}
D_l^{\rm odd}=\text{diag}\left(-l(l-1),-(l+2)(l+1)\right) \p
\end{equation}
We find that the two odd-parity eigen-solutions $(Q_l^{\rm odd})^{-1}\vec{R}^{\rm odd}$ decay as 
$r^{-l+1}$ and $r^{-l-1}$. Again we have a difference of two powers of $r$ in the decay rates, which is problematic. We cure this by taking $(Q_{l}^{\rm odd})_{i1}$ and $(M\omega)^2(Q_{l}^{\rm odd})_{i2}$
as our two initial amplitudes. 

For {\it even-parity} modes with $l\ge2$ we need to solve for the basis
$\vec{R} \equiv (R^{(1)}, R^{(3)}, R^{(5)}, R^{(6)}, R^{(7)})^T $ (the $l=0$ mode is obtained analytically---see Appendix \ref{App:monopole}). For $M\ll r\ll 1/\omega$, the ODEs again become analogous to Eq.\ \eqref{eq:large_r_R}, now with $A_1$ replaced with 
\begin{equation}
	A_l^{\rm even}= \left( \begin{array}{ccccc}
-(\Lambda+2) & 2 & 2 & 2 & 0 		\\
2 & -(\Lambda+2) & -2 & -2 & 0 	\\
2\Lambda & -2\Lambda & -(\Lambda+4) & -2\Lambda & 2	\\ 
2 & -2 & -2 & -(\Lambda+2) & 0	\\
0 & 0 & 2\Lambda-4 & 0 & -(\Lambda-2)\end{array} \right) \p
\end{equation}
We have
\begin{equation}
	Q_l^{\rm even}= \left( \begin{array}{ccccc}
\frac{1}{(l+2)(l+1)} & -\frac{1}{(l+2)(l-1)} & 1 & 1 & \frac{1}{l(l-1)} 		\\
-\frac{1}{(l+2)(l+1)} & 0 & 0 & 1 & -\frac{1}{l(l-1)}	\\
\frac{2}{l+2} & -\frac{1}{(l+2)(l-1)} & 0 & 0 & -\frac{2}{l-1}	\\ 
-\frac{1}{(l+2)(l+1)} & 0 & 1 & 0 & -\frac{1}{l(l-1)}	\\
1 & 1 & 0 & 0 & 1
\end{array} \right)
\end{equation}
and
\begin{equation}
D_l^{\rm even}=\text{diag}\left(-(l-1)(l-2),-\Lambda,-\Lambda,-\Lambda,-(l+2)(l+3)\right) \p
\end{equation}
We find that 3 of the eigen-solutions decay like $r^{-l}$, and the other two like $r^{-l-2}$ and $r^{-l+2}$, respectively. Here the problem is most acute, since the power-law variation is over {\it four} factors of $r$. We cure this by choosing our initial amplitudes to be
$(M\omega)^{-2}(Q_{l}^{\rm even})_{i1}$, $(Q_{l}^{\rm even})_{i2}$, $(Q_{l}^{\rm even})_{i3}$, $(Q_{l}^{\rm even})_{i4}$ and $(M\omega)^2(Q_{l}^{\rm even})_{i5}$.

\section{Sample results}\label{sec:results}

We present here a small sample of numerical GSF results from our code. Our sole purpose is to illustrate the correctness, accuracy and efficacy of our method---a physical application was presented in Ref.\ \cite{Warburton_etal}. The Lorenz-gauge GSF was calculated previously by Barack and Sago \cite{Barack-Sago-circular}, Berndtson \cite{Berndtson} and Akcay \cite{Akcay-GSF-circular} for circular orbits, and by Barack and Sago \cite{Barack-Sago-eccentric} for eccentric orbits (the latter using a time-domain method). We find results from our code to be in good agreement with published numerical data. 

Let us consider circular orbits first. Our code can take as input $e=0$ without modification, 
so this simple case already tests many of the program's routines. 
In Table \ref{table:circular_orbit_results} we show the radial component of the GSF for a sample of orbital radii $r_0$, alongside equivalent results from \cite{Barack-Sago-circular,Berndtson,Akcay-GSF-circular}. We find a good agreement as far out as $r_0=10000M$. 

\begin{table}[htb]
    \begin{center}
        \begin{tabular}{c || c | c  | c}
          \hline\hline
            $r_0/M$ & $(M/\mu)^2 F^r_\text{(this work)}$	& $(M/\mu)^2 F^r_\text{BS}$		&   $(M/\mu)^2 F^r_\text{B}$	\\
            \hline
						6			& $2.44664993(3)\times10^{-2}$		& $2.44661\times10^{-2}$			& $2.4466497\times10^{-2}$	\\
						10		& $1.33894695(7)\times10^{-2}$		& $1.33895\times10^{-2}$			& $1.3389470\times10^{-2}$	\\
						20		& $4.15705503(2)\times10^{-2}$		& $4.15706\times10^{-2}$			& $4.1570550\times10^{-2}$  \\
						50		& $7.44948594(7)\times10^{-4}$		& $7.44949\times10^{-4}$			& $7.4494860\times10^{-4}$	\\
						150		& $8.68274462(5)\times10^{-5}$		& $8.68274\times10^{-5}$			& $8.6827447\times10^{-5}$	\\
						500		& $7.9441064(8)\times10^{-6}$			& -								 						& $7.9441058\times10^{-6}$	\\
						800		& $3.1113443(3)\times10^{-6}$			& -								 						& $3.1113443\times10^{-6}$	\\
						10000	& $1.998(2)\times10^{-8}$					& -														& $1.9993000\times10^{-8}$	\\
			\hline
        \end{tabular}
\caption{Sample results for circular orbits: comparison with the literature.
The second column shows numerical values from our code for the radial component of the GSF. The third and fourth columns show equivalent results from Barack and Sago \cite{Barack-Sago-circular} ($F^r_\text{BS}$) and Berndtson \cite{Berndtson} ($F^r_\text{B}$). Entries were left empty where there is no published data available. A parenthetical figure indicates the estimated error in the last displayed decimal (Berndtson and Barack and Sago present only significant figures). \alert{The error bars we quote correspond to the difference in the GSFs computed using the inner and outer radial derivatives}.  Data points in column 2 were computed with $l_\max=50$ and took less than a minute each to produce on a 12 core 3GHz cluster ---a few hundred times less CPU time than the time-domain computation of \cite{Barack-Sago-circular}.}
	\label{table:circular_orbit_results}
    \end{center}
\end{table}
%
%

\begin{figure}
	\centering
	\includegraphics[width=85mm]{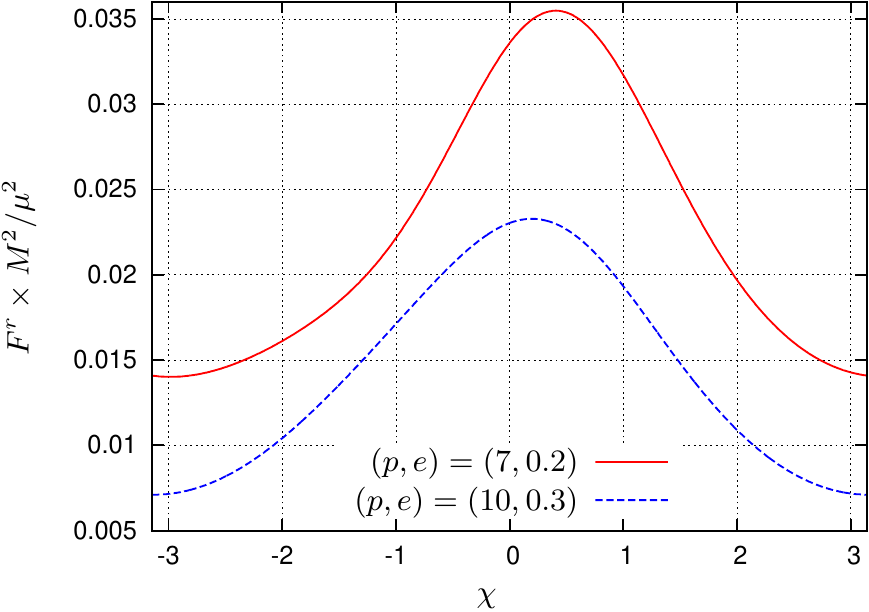}\hspace{0.5cm}
	\includegraphics[width=85mm]{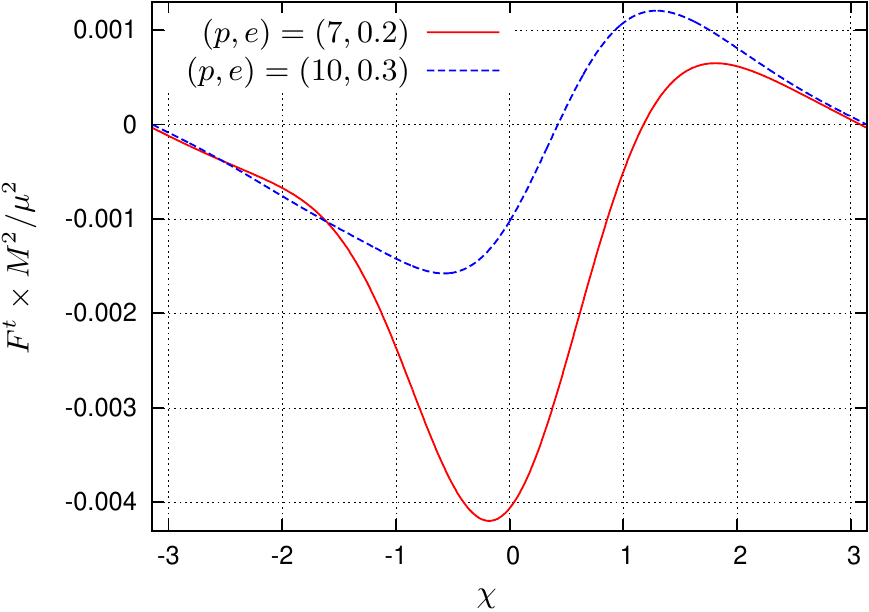}
	\caption{The total GSF ($r$ component on the left panel, $t$ component on the right) over a full radial period, for eccentric orbits with parameters $(p,e)=(7,0.2)$ and $(p,e)=(10,0.3)$. The radial phases $\chi=0$ and $\chi=\pm\pi$ correspond to periastron and apastron, respectively.}\label{fig:GSF_Fr_Ft}
\end{figure}

We next turn to eccentric orbits. Here the time-domain code by Barack and Sago \cite{Barack-Sago-eccentric} provides the only comparison point. In Tables \ref{table:GSF_p7e02} and \ref{table:GSF_p10e03} we present sample GSF data for two geodesic orbits, one with $(p,e)=(7, 0.2)$ and the other with $(p,e)=(10, 0.3)$, showing separately the conservative and dissipative pieces. Data for these orbits were also presented in Ref.\ \cite{Barack-Sago-eccentric}, and we find agreement \alert{in most cases to all significant figures}. Our results for these two orbits are about an order of magnitude more accurate than those presented in Ref.~\cite{Barack-Sago-eccentric}, taking an order of magnitude {\it less} CPU time to compute. In Fig.~\ref{fig:GSF_Fr_Ft} we plot $F^t$ and $F^r$ along the same two orbits, showing the variation of the GSF with the radial phase $\chi$.

\begin{table}[htb]
    \begin{center}
        \begin{tabular}{c || c | c | c | c}
          \hline\hline
			$\chi$		&	$(M/\mu)^2 F^t_\cons$		&	$(M/\mu)^2 F^t_\diss$		& $(M/\mu)^2 F^r_\cons$			& $(M/\mu)^2 F^r_\diss$													\\
			\hline
			0					&	0														&	$-4.0633017(3)\times10^{-3}$		&	$3.3576055(4)\times10^{-2}$		&	0															\\
			$\pi/4$		&	$8.64715(3)\times10^{-4}$		&	$-2.1569226(1)\times10^{-3}$		&	$2.9098813(5)\times10^{-2}$		&	$4.7349558(2)\times10^{-3}$		\\
			$\pi/2$		&	$8.286105(1)\times10^{-4}$	&	$-2.5168026(1)\times10^{-4}$		&	$2.1250343(6)\times10^{-2}$		&	$3.2041903(1)\times10^{-3}$		\\
			$3\pi/4$	&	$4.607495(2)\times10^{-4}$	&	$-1.1240916(2)\times10^{-5}$		&	$1.5901488(1)\times10^{-2}$		&	$9.6337335(3)\times10^{-4}$		\\
			$\pi$			&	0														&	$-3.4614164(4)\times10^{-5}$		&	$1.4088770(1)\times10^{-2}$		&	0															\\
			\hline
		\end{tabular}
		\caption[Sample GSF data for a geodesic orbit with parameters $(p,e)=(7, 0.2)$]{Sample GSF data for a geodesic orbit with parameters $(p,e)=(7, 0.2)$, showing separately the dissipative and conservative components. The GSF is sampled at a selection of radial phases $\chi$ along the orbit. The displayed error bars are computed by comparing results with $l_\max=15$ and $l_\max=20$. The $F^\varphi$ component can be constructed using the orthogonality relation $u_\alpha F^\alpha=0$, and recall $F^\theta=0$ by symmetry. \alert{The data presented in this table took approximately 12 minutes to generate on 64 cores of a cluster.}}\label{table:GSF_p7e02}
	\end{center}
\end{table}

\begin{table}[htb]
    \begin{center}
        \begin{tabular}{c || c | c | c | c}
          \hline\hline
			$\chi$		&	$(M/\mu)^2 F^t_\cons$		&	$(M/\mu)^2 F^t_\diss$		& $(M/\mu)^2 F^r_\cons$			& $(M/\mu)^2 F^r_\diss$		\\
			\hline
			0					&	0														&	$-1.0242488(1)\times10^{-3}$		&	$2.303161(2)\times10^{-2}$	&	0						\\
			$\pi/4$		&	$1.161566(4)\times10^{-3}$	&	$-3.6785582(2)\times10^{-4}$		&	$1.985394(1)\times10^{-2}$	&	$1.177853(1)\times10^{-3}$	\\
			$\pi/2$		&	$1.087278(2)\times10^{-3}$	&	$3.3433956(4)\times10^{-5}$			&	$1.362199(2)\times10^{-2}$	&	$5.654576(2)\times10^{-4}$	\\
			$3\pi/4$	&	$5.122832(1)\times10^{-4}$	&	$1.1041804(3)\times10^{-5}$			&	$8.810067(1)\times10^{-2}$	&	$1.0637516(1)\times10^{-4}$	\\
			$\pi$			&	0														&	$2.8361825(7)\times10^{-7}$			&	$7.110898(1)\times10^{-2}$	&	0						\\
			\hline
		\end{tabular}
		\caption[Sample GSF data for an orbit with parameters $(p,e)=(10 0.3)$]{Same as Table \ref{table:GSF_p7e02}, this time for $(p,e)=(10, 0.3)$. \alert{The displayed error bars are computed by comparing results with $l_\max=12$ and $l_\max=15$. The data presented in this table took approximately 24 minutes to generate on 64 cores of a cluster.} }\label{table:GSF_p10e03}
	\end{center}
\end{table}

\begin{figure}
	\centering
	\includegraphics[width=100mm]{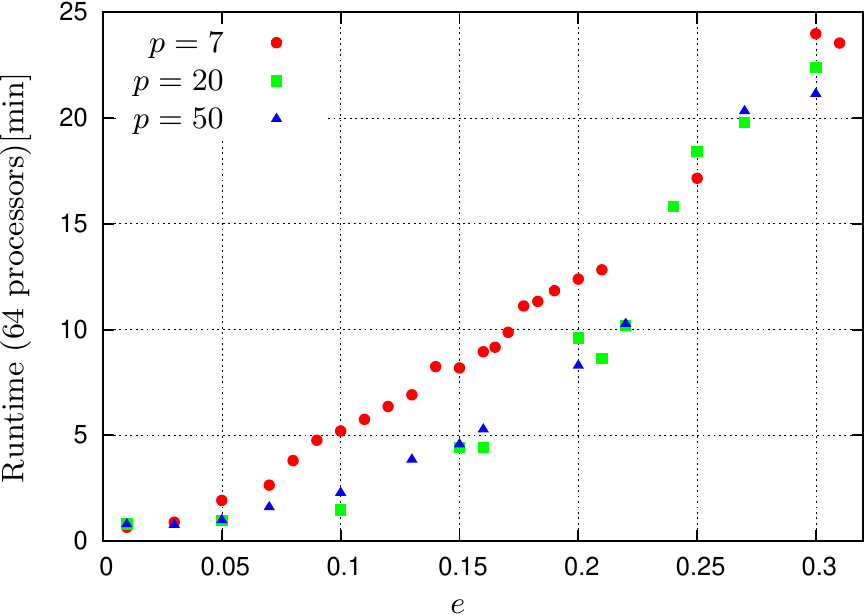}
	\caption[Computation time of FD GSF code on a cluster of 64 processors]
	{Computational performance. The plot shows the computation time for orbits with $p=\{7,20,50\}$ using a cluster of 64 processors with a target relative accuracy of $10^{-6}$ in the GSF. In each such ``computation'' all GSF components are obtained along an entire libration cycle of a fixed geodesic orbit with given $p,e$. 
	 }\label{fig:GSF_run_times}
\end{figure}

Figure \ref{fig:GSF_run_times} explores the computational performance of our code. The results illustrate the rapid increase in the computation burden with increasing eccentricity, as also found previously in calculations of the scalar-field self-force  \cite{Warburton-Barack:eccentric} and of the metric perturbation in the Regge--Wheeler gauge \cite{Hopper-Evans}. As discussed in Sec.\ \ref{sec:parallelization} our code is written to run on a computer cluster. By utilizing 64 processors, we are able to compute the GSF along an eccentric geodesic with given $e,p$ (at a relative accuracy of one part in $10^6$) in a matter of minutes---so long as the eccentricity is not larger than $\sim 0.3$. For eccentricities $< 0.05$ the computation takes under 2 minutes. As the eccentricity increases the runtime grows approximately linearly up to $e\sim0.2$. For higher eccentricities the runtime grows more rapidly. This increase in runtime is primarily due to the broadening of the Fourier spectrum as the orbital eccentricity increases. As an example, to reach our accuracy goal for $p = 7$, we require 6479 Fourier modes at $e = 0.1$, 9289 Fourier modes at $e = 0.2$, and as many as 15066 modes at $e = 0.3$. 

For orbits with $e\le0.2$, our code is capable of computing the GSF at 6-figure accuracy in under 12 hours on a standard (3GHz, dual core) desktop machine. This is an order of magnitude faster than comparable time-domain codes \cite{Barack-Sago-eccentric}. We have not obtained detailed timing data for eccentricities much above 0.3 (it becomes increasingly hard to avoid the low-$\omega$ problem at large eccentricities), but expect our code to remain competitive with time-domain implementations up to at least $e\sim 0.5$. This ability of our code to produce large quantities of data for low eccentricity orbits was a crucial prerequisite is enabling the orbital-evolution calculation presented in Ref.~\cite{Warburton_etal}.

\section{Outlook}\label{Sec:outlook}

We described here a computational framework for Lorenz-gauge GSF calculations in Schwarzschild spacetime. The framework allows efficient computations of the GSF along strong-field ($p\lesssim 50$) bound orbits of small and moderate eccentricities ($e\lesssim 0.3$). In the context of the ongoing GSF programme there are several high-priority improvements and extensions of our code, which we now briefly survey.

First, it is important to extend the reach of our method in parameter space, in two ways: farther out into the weak-field (large $p$) regime, and tighter around $\omega=0$ ``resonances''. The weak-field extension will allow interesting comparisons with post-Newtonian results and the calibration of post-Newtonian parameters using eccentric orbits, in much the same way as was done recently using circular orbits \cite{Blanchet-etal-PN-SF-comparision:circular,Blanchet_etal:PN_SF_comparison}. As a reference for comparison, one could utilize the eccentric-orbit generalization of Detweiler's ``redshift'' invariant \cite{Detweiler-circular}, proposed in Ref.\ \cite{Barack-Sago-precession}. The near-resonances extension is necessary to reduce/remove a troubling restriction on the range of strong-field orbits computable by our code. Both extensions require a better method for dealing with modes of very low frequency. Such a method is sketched in Ref.\ \cite{Warburton-thesis} but it is yet to be worked out in full and implemented. 

Second, it is a high-priority task to extend the computational framework to the Kerr case. There are several possible avenues of approach to this problem. The most straightforward (but lacking elegance and computationally tedious) would follow closely our treatment in Schwarzschild: One would write down the Lorenz-gauge perturbation equations on a Kerr background and decompose them into Fourier-harmonic modes using standard tensorial harmonic functions. The equations will couple between different $l$ multipoles (in addition to the usual coupling between tensorial components), and one would have to solve the resulting coupled set. Such an approach could prove either practical or not depending on the strength of the $l$-mode coupling. 

A second possibility (if one insists on working in the Lorenz gauge) is to derive a basis of ``tensorial spheroidal harmonics''---a generalization of both spin-2 spheroidal harmonics $_2S_{lm\omega}(\theta)$ \cite{Teukolsky} and tensorial spherical harmonics---that would fully separate the angular dependence in the linearized Einstein's equation (\ref{eq:linearized_einstein}) on a Kerr background. If this can be achieved, the Kerr problem could be tackled in much the same way as the Schwarzschild problem. (Such a useful basis of angular functions would have a wide range of applications in black hole perturbation theory beyond the GSF problem.)
 
A third option involves a departure from the pure-Lorenz-gauge strategy, as advocated in a series of papers by Friedman {\it et al.}~\cite{Keidl_etal:2006,Keidl-etal,Shah_etal,Shah_etal:Kerr} (and see also an early proposal in Ref.\ \cite{Barack-Ori-GSF_gauge_dependence}, and the recent Ref.\ \cite{Gralla:gauge_and_averaging}). The idea is to construct the GSF in some locally-regular gauge (i.e., a gauge related to Lorenz's via a sufficiently regular transformation), which is at the same time related to a {\it radiation} gauge \cite{Chrzanowski} via a simple, analytically given transformation. Since there is a known procedure for reconstructing the radiation-gauge perturbation from curvature scalars (solutions to the spin-2 Teukolsky equation) \cite{Chrzanowski,Ori:metric_reconstruction_in_Kerr}, we would reduce the numerical task to the solution of fully separable scalar-like equations. It is possible to make this idea work in practice by obtaining a modified version of the mode-sum formula (\ref{eq:ret-mode-sum}), in which the numerical input is (essentially) modes of the radiation-gauge metric perturbation (and their derivatives), and the regularization parameters are modified to account for the local gauge transformation to Lorenz gauge. The details of this method, which we consider a most promising approach to the Kerr problem, will be presented in a forthcoming paper \cite{Barack-Merlin-Ori-Pound}.

Third, as an even more ambitious extension of our method, consider the problem of computing {\it second-order} GSF effects. Several formulations of the GSF at second order in the mass ratio have been presented recently \cite{Detweiler:2nd_order,Pound:2nd_order,Gralla:2nd_order,Pound:2nd_order2}. Ref.\ \cite{Pound:2nd_order2}, in particular, proposes a practical method for computing the regularized second-order metric perturbation; this can be used, e.g., to compute Detweiler's redshift variable through second-order in the mass ratio. At the practical level, the numerical task reduces to the solution of the linearized Einstein's equation (\ref{eq:linearized_einstein}), sourced by a certain extended ``effective source'' constructed from the (regularized) first-order perturbation and its derivatives, as prescribed in \cite{Pound:2nd_order2}. This problem can be tackled using a suitably modified version of our existing FD code. Work is in progress to implement this method numerically for circular orbits around a Schwarzschild black hole. 

\section*{Acknowledgements}

SA and LB acknowledge support from STFC through grant number PP/E001025/1. NW's work was supported by the Irish Research Council, which is funded under the National Development Plan for Ireland. The research leading to these results received funding from the European Research Council under the European Union's Seventh Framework Programme (FP7/2007-2013)/ERC grant agreement no. 304978. 

\appendix

\section{Coupling terms in the field equations, and source functions}\label{apdx:field_eqs_and_sources}
We give here the explicit form of the terms $\hat{\mathcal{M}}^{(i)l}_{(j)}$ and $J^{(i)}_{lmn}$ appearing in the FD field equations (\ref{eq:metric_pert_FD}). Omitting the modal indices $ lmn $ for simplicity, the $\hat{\mathcal{M}}^{(i)l}_{(j)}$ are given by
\ba
\hat{\mathcal{M}}^{(1)}_{(j)} R^{(j)} & = &  \f{M}{r^2} f R^{(3)}_{,r_\ast} + \f{f}{2r^2}\left(1-\f{4M}{r}\right) \left(R^{(1)}-R^{(5)} - f R^{(3)} \right) - \f{f^2}{2r^2} \left(1-\f{6M}{r}\right) R^{(6)}, \label{eq:eq_R1} \\
\hat{\mathcal{M}}^{(2)}_{(j)} R^{(j)} & = &  \f{1}{2}ff' R^{(3)}_{,r_\ast} + \f{1}{2} f'  \left[ i \omega \left(R^{(1)}-R^{(2)}\right) + R^{(2)}_{,r_\ast} - R^{(1)}_{,r_\ast} \right] + \f{f^2}{2 r^2} \left(R^{(2)}-R^{(4)}\right) \nn \\ & \quad & \qquad + \f{f f'}{2r}\left( R^{(1)} - R^{(5)} - f R^{(3)} - 2f R^{(6)} \right), \label{eq:eq_R2} \\
\hat{\mathcal{M}}^{(3)}_{(j)} R^{(j)} & = &  - \f{f}{2r^2} \left[R^{(1)} - R^{(5)} - \left(1-\f{4M}{r}\right) \left(R^{(3)} + R^{(6)}\right) \right], \label{eq:eq_R3} \\
\hat{\mathcal{M}}^{(4)}_{(j)} R^{(j)} & = &  \f{1}{4} f' \left[ i \omega \left(R^{(5)}-R^{(4)}\right) + R^{(4)}_{,r_\ast} - R^{(5)}_{,r_\ast} \right] - \f{1}{2} l (l+1) \f{f}{ r^2} R^{(2)} \nn \\ &\quad& \qquad - \f{f f'}{4r}\left( 3R^{(4)} + 2R^{(5)} - R^{(7)} + l (l+1) R^{(6)} \right)  , \label{eq:eq_R4} \\
\hat{\mathcal{M}}^{(5)}_{(j)} R^{(j)} & = &  \f{f}{r^2} \left[ \left(1-\f{9M}{2r}\right) R^{(5)} - \f{l(l+1)}{2}\left(R^{(1)} - f R^{(3)} \right) + \f{1}{2} \left(1-\f{3M}{r}\right) \left( l(l+1) R^{(6)} - R^{(7)} \right) \right], \label{eq:eq_R5} \\
\hat{\mathcal{M}}^{(6)}_{(j)} R^{(j)} & = &  - \f{f}{2r^2} \left[R^{(1)} - R^{(5)} - \left(1-\f{4M}{r}\right) \left(R^{(3)} + R^{(6)}\right) \right], \label{eq:eq_R6} \\
\hat{\mathcal{M}}^{(7)}_{(j)} R^{(j)} & = &  - \f{f}{2r^2} \left(R^{(7)} + \lambda R^{(5)} \right), \label{eq:eq_R7} \\
\hat{\mathcal{M}}^{(8)}_{(j)} R^{(j)} & = & \f{1}{4} f' \left[ i \omega \left(R^{(9)}-R^{(8)}\right) + R^{(8)}_{,r_\ast} - R^{(9)}_{,r_\ast} \right] -  \f{f f'}{4r}\left( 3R^{(8)} + 2R^{(9)} - R^{(10)}  \right) , \label{eq:eq_R8} \\
\hat{\mathcal{M}}^{(9)}_{(j)} R^{(j)} & =&  \f{f}{r^2}\left(1-\f{9M}{2r}\right) R^{(9)} - \f{f}{2r^2}\left(1-\f{3M}{r}\right) R^{(10)}, \label{eq:ODE_9} \\  
\hat{\mathcal{M}}^{(10)}_{(j)} R^{(j)} & =&   -\f{f}{2r^2} R^{(10)} - \f{f \lambda}{2r^2} R^{(9)},  \label{eq:ODE_10}
\ea
where, recall, $ f = 1-2M/r,\: f'= \partial f/\partial r $ and $ \lambda = (l+2)(l-1) $.

The FD source functions $ J^{(i)}_{lmn} \equiv  J^{(i)}_{lmn}(r) $ appearing in Eq.~(\ref{eq:metric_pert_FD}) are derived from the time-domain source functions $\mathcal{S}^{(i)}_{lm}\equiv\mathcal{S}^{(i)}_{lm}(t,r)$ of Eq.~\eqref{eq:metric_pert_1+1} using
\be
J^{(i)}_{lmn} = -\frac{4}{T_r}\int^{T_r/2}_{-T_r/2} \mathcal{S}^{(i)}_{lm}\delta(r-r_p) e^{i\omega t}\, dt \c \label{eq:J_i_app}
\ee
where $T_r$ is the radial period of Eq.~(\ref{eq:T_r}). The functions $\mathcal{S}^{(i)}_{lm}$ themselves are given by \cite{Barack-Sago-eccentric}
%
%
%
%
%

\begin{align}
	\mathcal{S}^{(1)}_{lm} 	&=	\mu\frac{4\pi f^2_p}{\en r_p^3}(2\en^2r_p^2-f_p r_p^2-\ang^2 f_p)Y^*_{lm}(\pi/2,\varphi_p)		\c		\\
	\mathcal{S}^{(2)}_{lm} 	&=	-\mu\frac{8\pi f_p^2}{r_p} u^r Y^*_{lm}(\pi/2,\varphi_p)											\c	\\
	\mathcal{S}^{(3)}_{lm} 	&=	\mu\frac{4\pi}{\en r_p^3} f^2_p (r_p^2 + \ang^2)Y^*_{lm}(\pi/2,\varphi_p)							\c	\\
	\mathcal{S}^{(4)}_{lm} 	&=	\mu\frac{8\pi im f^2_p \ang}{r_p^2} Y^*_{lm}(\pi/2,\varphi_p)										\c	\\
	\mathcal{S}^{(5)}_{lm} 	&=	-\mu\frac{8\pi im f_p^2 u^r \ang}{r_p^2 \en} Y^*_{lm}(\pi/2,\varphi_p)					\label{S5}			\c	\\
	\mathcal{S}^{(6)}_{lm} 	&=	\mu\frac{4\pi f_p^2 \ang^2}{r_p^3\en}Y^*_{lm}(\pi/2,\varphi_p)										\c	\\
	\mathcal{S}^{(7)}_{lm} 	&=	\left[l(l+1)-2m^2\right] \mathcal{S}^{(6)}_{lm}													\c	\\
	\mathcal{S}^{(8)}_{lm} 	&=	\mu\frac{8\pi f_p^2 \ang}{r_p^2} Y^*_{lm,\theta}(\pi/2,\varphi_p)							\c			\\
	\mathcal{S}^{(9)}_{lm} 	&=	-\mu\frac{8\pi f^2_p u^r \ang}{r_p^2\en}Y^*_{lm,\theta}(\pi/2,\varphi_p)							\c	\\
	\mathcal{S}^{(10)}_{lm}  &=	\mu\frac{8\pi imf^2_p \ang^2}{r_p^3\en}Y^*_{lm,\theta}(\pi/2,\varphi_p) \p
\end{align}

Here, recall that the subscript `p' denotes the value of a quantity at the particle and $\en$ and $\ang$ are given by Eq.~\eqref{eq:Schwarzschild_Energy_Ang_Mom}.

The integral in Eq.~(\ref{eq:J_i_app}) is readily evaluated. For example, for $i=5$ we have
\begin{eqnarray}
	J^{(5)}_{lmn} &= &-\frac{4}{T_r}\int^{T_r/2}_{-T_r/2} \mathcal{S}^{(5)}_{lm}\delta(r-r_p) e^{i\omega t}\, dt  \nonumber\\
		&=& \mu\frac{32\pi i m {\cal Y}_{lm}}{T_r\en}\int^{T_r/2}_{-T_r/2} \frac{f_p^2 u^r}{r_p^2} e^{i(\omega t-m\varphi_p)}\delta(r-r_p)\, dt		\nonumber\\
	&=& -\mu\frac{64\pi m {\cal Y}_{lm}}{T_r\en}\int^{T_r/2}_0 \frac{f_p^2 u^r}{r_p^2} \sin(\omega t - m\varphi_p)\delta(r-r_p)\, dt ,
\end{eqnarray}
where we have introduced $ {\cal Y}_{lm} \equiv Y_{lm}(\pi/2, 0) $. In moving from the first line to the second we have substituted from Eq.\ (\ref{S5}), leaving inside the integral all $t$-dependent quantities [like $f_p=1-2M/r_p(t)$]. In moving to the third line, we have made use of the orbital symmetries: taking $t=\varphi=0$ at some periastron passage, we have that $r_p(t)$ is an even function while $u^r(t)$ and $\varphi_p(t)$ are odd, and it follows that the real part of the integrand [$\propto \cos(\omega t-m\varphi)$] is an odd function of $t$ while the imaginary part [$\propto i\sin(\omega t-m\varphi)$] is even. Performing the integration we finally obtain 
\begin{equation}
	J^{(5)}_{lmn} = -\mu\frac{64\pi m  u^t f_p^2  \ang}{T_r \en r_p^2} {\cal Y}_{lm} \sin(\omega_{mn}t_p - m\varphi_p) \p
\end{equation}

The other FD source functions are evaluated in a similar fashion. We obtain
\begin{align}
	J^{(1)}_{lmn} &= -\mu\frac{32\pi u^t f^2_p}{T_r \en r_p^3 |u^r|} (2\en^2r_p^2 - fr_p^2 - \ang^2f_p) {\cal Y}_{lm} \cos(\omega_{mn}t_p - m\varphi_p) \label{eq:J1}		\c \\
	J^{(2)}_{lmn} &= \mu\frac{64\pi i u^t f_p^2}{T_r r_p} {\cal Y}_{lm} \sin(\omega t - m\varphi_p) \c		\\
	J^{(3)}_{lmn} &= -\mu\frac{32\pi u^t f^2_p}{T_r \en r_p^3 |u^r|} (r_p^2 + \ang^2) {\cal Y}_{lm}	\cos(\omega_{mn}t_p - m\varphi_p) \label{eq:J3}				\c	\\
	J^{(4)}_{lmn} &= -\mu\frac{64 i m u^t f_p^2 \ang}{T_r r_p^2 |u^r|} {\cal Y}_{lm} \cos(\omega t - m\varphi_p)		\c \\
	J^{(6)}_{lmn} &= -\mu\frac{32\pi u^t f_p^2 \ang^2}{T_r \en r_p^3 |u^r|} {\cal Y}_{lm} \cos(\omega_{mn}t_p - m\varphi_p)						\c	\\
	J^{(7)}_{lmn} &= \mu\left[l(l+1)-2m^2\right]J^{(6)_{lmn}}		\c	\\
	J^{(8)}_{lmn} &= -\mu\frac{64\pi  u^t \ang f_p^2}{T_r  r_p^2 |u^r|} {\cal Y}_{lm,\theta} \cos(\omega t - m\varphi_p)		\c	\\
	J^{(9)}_{lmn} &= \mu\frac{64 i\pi f_p^2  u^t \ang}{T_r \en r_p^2}  {\cal Y}_{lm,\theta} \sin(\omega_{mn}t_p - m\varphi_p)		 	\c				\\
	J^{(10)}_{lmn} &= -\mu\frac{64i\pi m u^t \ang^2 f_p^2}{T_r \en r_p^3 |u^r|} {\cal Y}_{lm,\theta} \cos(\omega_{mn}t_p - m\varphi_p)			\c
\end{align}
where $ {\cal Y}_{lm,\theta}\equiv Y_{lm,\theta}(\pi/2,0) $.

\section{Static piece of the monopole mode (the ``mass perturbation'')}\label{App:monopole}

The static, spherically symmetric mode of the metric perturbation, $l=m=n=0$, is, up to a gauge piece, simply a mass variation of the background Schwarzschild geometry, caused by the mass-energy $\mu{\cal E}$ of the particle.
Detweiler and Poisson \cite{Detweiler-Poisson} derived this mode analytically, in the Lorenz gauge, for a particle in circular geodesic motion (see \cite{Barack-Lousto-2005} for an explicit expression). 
Barack and Sago \cite{Barack-Sago-eccentric} later generalized this to eccentric orbits, but their report did not contain full details of the calculation. Here we take the opportunity to complete these details, while also demonstrating the application of MEHS.


For $l=m=0$, the FD Lorenz-gauge field equations \eqref{eq:metric_pert_FD} reduce to three equations, for $i=1,3,6$; the other $i$-modes vanish identically. Of the four FD gauge conditions (\ref{eq:gauge1})-(\ref{eq:gauge4}), the only nontrivial is (\ref{eq:gauge2}), which expresses $R^{(6)}$ algebraically in terms of $R^{(1)}$ and $R^{(3)}$ (and their radial derivatives). The system thus reduces to a set of two coupled second-order ODEs.  Therefore, the complete basis of homogeneous static monopole solutions (in the Lorenz gauge) is {\em four-dimensional}.
 
Let us denote by 
\begin{eqnarray}
H &\equiv & (M/\mu)\left\{h_{tt},\ h_{rr},\ r^{-2}h_{\theta\theta}= (r\sin\theta)^{-2}h_{\varphi\varphi}\right\}
\nonumber\\
&=& 
\frac{M}{4\sqrt{\pi}r}\left\{ R^{(1)}+fR^{(6)},\ f^{-2}(R^{(1)}-fR^{(6)}),\ R^{(3)}\right\} 
\end{eqnarray}
the metric perturbation corresponding to a homogeneous solution
$R^{(i)}=\left\{ R^{(1)}, R^{(3)},R^{(6)}\right\}$ to the set \eqref{eq:metric_pert_FD} for $l=m=n=0$ (other components of the perturbation vanish). The inverse relations are 
\ba
R^{(1)} & = & 2 \sqrt{\pi} \mu^{-1} r \left(h_{tt} + f^2 h_{rr} \right) \label{eq:h1_mono}, \\
R^{(3)} & = & 4 \sqrt{\pi} \mu^{-1} r^{-1} h_{\theta\theta} \label{eq:h3_mono}, \\
R^{(6)} & = & 2 \sqrt{\pi} \mu^{-1} \f{r}{f} \left(h_{tt} - f^2 h_{rr} \right). \label{eq:h6_mono}
\ea
In terms of $H$, a {\em complete} basis of linearly independent homogeneous solutions is given by 
\footnote{Ref.\ \cite{Dolan-Barack:GSF_m_mode} gives an alternative basis of homogeneous solutions, with $H_{A,B,C}$ as here (up to normalization), but with a solution $H_{D}$ that tends to a constant at infinity. The basis in \cite{Dolan-Barack:GSF_m_mode} is constructed so that $H_{B,C,D}$ span the 3-dimensional subspace of pure-gauge static monopole perturbations. In our choice, which is differently motivated, $H_{D}$ is not pure gauge; see below.} 
\begin{mathletters}
\begin{equation}\label{general1}
H_A =
\left\{-f,f^{-1},1\right\},
\end{equation}
\begin{equation}\label{general2}
H_B=
\left\{-\frac{fM}{r^3}P(r),\frac{f^{-1}}{r^3}Q(r),\frac{f}{r^2} P(r)\right\},
\end{equation}
\begin{equation}\label{general3}
H_C=
\left\{-\frac{M^4}{r^4},\frac{M^3f^{-2}(3M-2r)}{r^4},\frac{M^3}{r^3}\right\},
\end{equation}
\begin{eqnarray}\label{general4}
H_D&=&
\left\{\frac{M}{r^4}\left[W(r)+rP(r)f\ln f-8M^3\ln(r/M)\right],
\right.\nonumber\\ &&
\frac{f^{-2}}{r^4}\left[K(r)-r Q(r) f\ln f-8M^3(2r-3M)\ln(r/M)
\right],
\nonumber\\ && \left.
\frac{1}{r^3}\left[3r^3-W(r)-rP(r)f\ln f+8M^3\ln(r/M)\right]\right\},
\end{eqnarray}
 \end{mathletters}
where $P,Q,W,K$ are polynomials in $r$:
\begin{eqnarray}
P(r)&=& r^2+2rM+4M^2,            \nonumber\\
Q(r)&=& r^3-r^2M-2rM^2+12M^3,    \nonumber\\
W(r)&=&3r^3-r^2M-4rM^2-28M^3/3,  \nonumber\\
K(r)&=&r^3M-5r^2M^2-20rM^3/3+28M^4. 
\end{eqnarray}

The solutions $H_A$ and $H_B$ are regular at the event horizon (as can be checked by moving to horizon-regular coordinates) but they fail to fall off at $r\to\infty$, instead approaching finite nonzero values. On the other hand, the solutions $H_C$ and $H_D$ are regular at infinity, but they are singular on the horizon. Therefore, it is tempting to select $\{H_A,H_B\}$ as our ``internal'' pair of basis functions, and take $\{H_C,H_D\}$ as our ``external'' pair.  We would then take 
\begin{eqnarray}\label{eq:monopole_gen_inh_sol}
\tilde H_{-}&=& C_A H_A + C_B H_B \c \nonumber\\
\tilde H_{+}&=& C_C H_C + C_D H_D
\end{eqnarray}
as our internal and external extended homogeneous solutions, with coefficients $C_j$ to be determined as prescribed in Eq.\ (\ref{eq:weighting_coeffs2}), i.e.,
\be
( C_A\ C_B\ C_C\ C_D)^T = \int_0^\pi \Phi^{-1} ( 0 \ 0 \ \hat J^{(1)} \ \hat J^{(3)} )^T \,\frac{d\tau}{dt}\frac{dt}{d\chi} \, f^{-1} d\chi, \label{eq:mono_matrix_eq}
\ee
with
\be
\Phi = \left(\begin{array}{cccc} -R^{(1)}_A & -R^{(1)}_B & R^{(1)}_C & R^{(1)}_D \\  -R^{(3)}_A & -R^{(3)}_B & R^{(3)}_C & R^{(3)}_D \\  -R^{(1)}_{A,{r_*}} & -R^{(1)}_{B,{r_*}} & R^{(1)}_{C,{r_*}} & R^{(1)}_{D,{r_*}} \\  -R^{(3)}_{A,{r_*}} & -R^{(3)}_{B,{r_*}} & R^{(3)}_{C,{r_*}} & R^{(3)}_{D,{r_*}}  \end{array}\right) \c \label{eq:mono_matrix}
\ee
where we denoted the $R^{(i)}$ functions corresponding to $H_A,\ldots,H_D$ by $R^{(i)}_A,\ldots,R^{(i)}_D$.
%
%
In the vacuum outside the libration region, the homogeneous solutions $\tilde H_{\pm}$ coincide with a particular {\em inhomogeneous} solution to the $l=m=n=0$ equations---call it $H_{\rm ih}$---such that
$H_{\rm ih}=\tilde H_{-}$ for $2M<r<r_{\min}$ and $H_{\rm ih}=\tilde H_{+}$ for $r>r_{\max}$. As we now explain, the solution $H_{\rm ih}$ associated with the choice (\ref{eq:monopole_gen_inh_sol}) of EHS {\em does not represent a physical mass perturbation}, which makes the choice (\ref{eq:monopole_gen_inh_sol}) physically inappropriate.

To see this, let us first recall that, if the perturbation $H_{\rm ih}$ is to represent a physical mass perturbation, then (i) in the vacuum region $2M<r<r_{\min}$ it must be pure gauge, and (ii) in the vacuum region $r>r_{\max}$ it must be given (up to a gauge piece) by a mass variation of the Schwarzschild background with an amplitude of precisely $\mu{\cal E}$. Ref.\ \cite{Dolan-Barack:GSF_m_mode} discusses a practical way to ``measure'' the gauge-invariant mass-energy content of a given perturbation, and in particular it is found that the solution $H_A$ represents a mass perturbation with mass-energy $1/2$, while $H_B$ is pure gauge and carries no mass-energy. It follows that the solution $H_A$ must not feature in $2M<r<r_{\min}$. However, in practice one generally finds $C_A\ne 0$, which implies that our EHS $\tilde H_{-}$ does feature the mass-full mode $H_A$ in the interior. Hence, the corresponding inhomogeneous solution $H_{\rm ih}$ is not pure gauge in $2M<r<r_{\min}$ and cannot represent a physical mass perturbation. 

The difference between the non-physical solution $H_{\rm ih}$ and the desired physical mass-perturbation solution---call it $H_{\rm ih}^{\delta M}$---must be given as a certain linear combination of homogeneous solutions $H_A,\ldots,H_D$. We now aim to find this ``mass fixing'' correction, 
\begin{equation}
\Delta H_{\rm ih}\equiv H_{\rm ih}^{\delta M}-H_{\rm ih} \p
\end{equation}  
First, we note that $\Delta H_{\rm ih}$ cannot contain $H_C$ or $H_D$, since both these basis functions are singular at the event horizon (and note that no linear combination of $H_C$ or $H_D$ is horizon-regular). Of the two remaining solutions, only $H_A$ possesses mass-energy, so the only way to guarantee that the mass-energy in the region $2M<r<r_{\min}$ vanishes is to subtract $C_A H_A$ off $H_{\rm ih}$. However, this introduces an irregularity at infinity---which can only be regulated by adding a suitable multiple of $H_B$. Noting the asymptotic forms $H_A\to\{-1,1,1\}$ and $H_B\to\{0,1,1\}$ as $r\to\infty$, we see that we may at best achieve ``asymptotic flatness'' in the spatial part of the metric, by making  the choice 
\begin{equation}
\Delta H_{\rm ih}=-C_A (H_A-H_B) \p
\end{equation}
This results in a perturbation $H_{\rm ih}^{\delta M}$ whose $tt$ component approaches a nonzero constant (=$C_A$) at infinity. This minor peculiarity of the Lorenz gauge is well known \cite{Barack-Lousto-2005}, and it is easily remedied with a simple rescaling of the $t$ coordinate [formally an $\mathcal{O}(\mu)$ gauge transformation away from Lorenz gauge], as discussed in \cite{Sago-Barack-Detweiler,Damour-EOB-SF,Dolan-Barack:GSF_m_mode} (for circular orbits) and in \cite{Barack-Sago-eccentric} (for eccentric orbits).

The above construction yields 
\begin{equation} \label{eq:physical_mass_perturbation}
	H_{\rm ih}^{\delta M}= H_{\rm ih} +\Delta H_{\rm ih}= \left\{
     \begin{array}{lr}
      (C_A+C_B) H_B , &	2M<r<r_{\min},\\
       -C_A (H_A-H_B) + C_C H_C + C_D H_D ,	& r > r_\max.
     \end{array}
   \right.
\end{equation}
It remains to verify that the perturbation $H_{\rm ih}^{\delta M}$ has the correct mass-energy content (i.e., $\mu{\cal E}$) in the domain $r>r_{\max}$. Since $H_A$ has mass-energy of $1/2$ and $H_D$ has mass-energy of $3/2$ ($H_B$ and $H_C$ are pure gauge), we need 
\be
\frac{3}{2}C_D-\frac{1}{2}C_A= \mu{\cal E} \label{eq:mono_energy} .
\ee
We prove this relation by direct calculation, as follows. From Eq.\ (\ref{eq:mono_matrix_eq}) we have 
\be
\frac{3}{2}C_D-\frac{1}{2}C_A=
\int_{t(\chi=0)}^{t(\chi=\pi)} \f{1}{2}\left[ \left( 3\Phi^{-1}_{43} - \Phi^{-1}_{13}\right) \hat J^{(1)} + (3\Phi^{-1}_{44}-\Phi^{-1}_{14}) \hat J^{(3)} \right] \f{d\tau}{dt} f^{-1} dt, \label{eq:mono_energy2}
\ee
%
where the relevant elements of $\Phi^{-1}$ work out as
\be 
\Phi^{-1}_{13}= \frac{1}{4f^2\pi^{1/2}},\quad\quad
\Phi^{-1}_{14}=0, \quad\quad
\Phi^{-1}_{43}=\frac{4-r/M}{24f^2\pi^{1/2}}, \quad\quad
\Phi^{-1}_{44}=-\frac{(r/M)}{24\pi^{1/2}},\quad\quad
\ee
and the source terms $\hat J^{(1,3)}$ (or rather $J^{(1,3)}= \hat J^{(1,3)}/u^r$) are given in Eqs.\ (\ref{eq:J1}) and (\ref{eq:J3}). A line of algebra then shows that the integrand in Eq.\ (\ref{eq:mono_energy2}) reduces to $2\mu{\cal E}/T_r$, and Eq.\ (\ref{eq:mono_energy}) follows by virtue of 
$2\int_{t(\chi=0)}^{t(\chi=\pi)} dt =T_r$.

%
%

%

We have thereby constructed a (unique) physical mass perturbation in the Lorenz gauge. The form of this perturbation in the vacuum region outside the libration domain is described in Eq.\ (\ref{eq:physical_mass_perturbation}). Our extended homogeneous solutions must coincide with this perturbation in the vacuum region. Hence, we must take (reverting to the $R^{(i)}$ notation, as in the main text) 
\begin{eqnarray}\label{eq:monopole_gen_inh_sol2}
\tilde R_{-}^{(i)}&=& (C_A+C_B) R^{(i)}_B \c \nonumber\\
\tilde R_{+}^{(i)}&=& -C_A (R^{(i)}_A-R^{(i)}_B) + C_C R^{(i)}_C + C_D R^{(i)}_D \c
\end{eqnarray}
where, recall, the coefficients $C_A,\ldots,C_D$ are those calculated using Eq.\ (\ref{eq:mono_matrix_eq}).

In summary, to construct the EHS for the mode $l=m=n=0$, one starts with the analytic solutions (\ref{general1})--(\ref{general4}), and from them constructs the corresponding functions $R^{(i)}_A,\ldots,R^{(i)}_D$ via Eqs.\ (\ref{eq:h1_mono})--(\ref{eq:h6_mono}). One then constructs the matrix $\Phi$ and computes the coefficients $C_A,\ldots,C_D$ using Eq.\ (\ref{eq:mono_matrix_eq}). The desired EHS are then given by Eqs.\ (\ref{eq:monopole_gen_inh_sol2}).

\bibliographystyle{apsrev4-1}
\bibliography{references}

\end{document}